\documentclass[12pt,a4paper]{article}

\usepackage[intlimits]{amsmath}
\usepackage{amssymb}
\usepackage{axodraw}
\usepackage{bbm}
\usepackage{enumerate}
\usepackage{cite} % [1-5] instead of [1,2,3,4,5]
\usepackage{subfigure}
\usepackage[scale={.75,.75}]{geometry}
\usepackage{graphicx}
\usepackage{caption}
\def\clap#1{\hbox to 0pt{\hss#1\hss}}
\def\mathclap{\mathpalette\mathclapinternal}
\def\mathrlap{\mathpalette\mathrlapinternal}

\def\mathclapinternal#1#2{\clap{$\mathsurround=0pt#1{#2}$}}
\def\mathrlapinternal#1#2{\rlap{$\mathsurround=0pt#1{#2}$}}

\DeclareMathOperator{\im}{Im}

\DeclareMathOperator{\tr}{tr}

\DeclareMathOperator{\diag}{diag}
\renewcommand{\d}{\text{d}}

\newcommand{\I}{\text{i}}

\newcommand{\gt}{\ensuremath{\tilde{g}}}
\newcommand{\pt}{\ensuremath{\tilde{p}}}
\newcommand{\qt}{\ensuremath{\tilde{q}}}
\newcommand{\et}{\ensuremath{\tilde{e}}}
\newcommand{\vt}{\ensuremath{\tilde{v}}}
\newcommand{\ut}{\ensuremath{\tilde{u}}}
\newcommand{\wt}{\ensuremath{\tilde{w}}}

\newcommand{\rhot}{\ensuremath{\tilde{\rho}}}

\newcommand{\omegat}{\ensuremath{\widetilde{\omega}}}

\newcommand{\Mt}{\ensuremath{\widetilde{M}}}
\newcommand{\Pt}{\ensuremath{\widetilde{P}}}
\newcommand{\Rt}{\ensuremath{\widetilde{R}}}
\newcommand{\Sigmat}{{\ensuremath{\widetilde{\Sigma}}}}
\newcommand{\Kt}{{\widetilde{K3}}}
\newcommand{\Vt}{\widetilde{V}}
\newcommand{\Ct}{\widetilde{C}}
\newcommand{\nut}{\widetilde{\nu}}
\newcommand{\jt}{\widetilde{j}}
\newcommand{\etat}{\widetilde{\eta}}
\newcommand{\Vv}{\mathcal{V}}
\newcommand{\Nn}{\mathcal{N}}

\newcommand{\Pp}{\mathbb{P}}
\newcommand{\Ppt}{\widetilde{\mathbb{P}}}

\numberwithin{equation}{section}
\numberwithin{table}{section}
\allowdisplaybreaks

\begin{document}
\date{\mbox{ }}

\title{ 
{\normalsize     
HD-THEP-08-27\hfill\mbox{}\\
%November 2008\hfill\mbox{}\\
}
\vspace{1cm}
\bf 
Fixing D7 Brane Positions\\by F-Theory Fluxes\\[8mm]}
%
%\vspace{2cm} 
\author{A.~P.~Braun$^1$, A.~Hebecker$^1$, C.~L\"udeling$^{1,2}$, R.~Valandro$^1$\\[2mm]
{\normalsize\itshape  $^1$~Institut f\"ur Theoretische Physik, Universit\"at Heidelberg,}\\
{\normalsize\itshape Philosophenweg 16-19, 69120 Heidelberg, Germany}\\
{\normalsize\itshape  $^2$~Bethe Center for Theoretical Physics and
Physikalisches Institut der Universit\"at Bonn,}\\
{\normalsize\itshape Nussallee 12, 53115 Bonn, Germany}\\[2mm]
{\normalsize {\ttfamily a.braun}, {\ttfamily a.hebecker}, {\ttfamily r.valandro} {\ttfamily
    @thphys.uni-heidelberg.de}, }\\ 
{\normalsize \ttfamily luedeling@th.physik.uni-bonn.de}}
%\date{Version: \today}
\maketitle

\thispagestyle{empty}

\vspace{0.7cm}
\begin{abstract}
\noindent

To do realistic model building in type IIB supergravity, it is important to
understand how to fix D7-brane positions by the choice of fluxes. More
generally, F-theory model building requires the understanding of how fluxes
determine the singularity structure (and hence gauge group and matter content)
of the compactification. We analyse this problem in the simple setting
of M-theory on $K3\times K3$. Given a certain flux which is consistent with
the F-theory limit, we can explicitly derive the positions at which D7~branes
or stacks of D7~branes are stabilised. The analysis is based on a parameterization
of the moduli space of type IIB string theory on $T^2/\mathbb{Z}_2$ (including
D7-brane positions) in terms of the periods of integral cycles of M-theory on
$K3$. This allows us, in particular, to select a specific desired gauge
group by the choice of flux numbers.

\end{abstract}

\newpage
\tableofcontents
\newpage

\section{Introduction\label{sec:intro}}

Over the past years, significant progress in some of the central 
phenomenological problems of string theory compactifications has been made. These 
problems include, in particular, moduli stabilisation, SUSY breaking, inflation and 
the possibility of fine-tuning the cosmological constant in the landscape of 
flux vacua. Much of this has been realised most successfully and explicitly 
in type IIB Calabi--Yau orientifold models in the supergravity
regime (for reviews, see e.g.~\cite{g05,dk06,bkl06,d08,ms07}). 
However, obtaining the standard model particle spectrum remains difficult in 
this context, although successful local constructions exist\cite{bhv08,dw08,cmq03,vw05,ms04,wy04}.

Motivated by the desire to make progress towards a type IIB (or, more 
generally, F-theory) derivation of the standard model, the present paper
analyses the way in which fluxes determine D7 brane positions. This is 
central to weakly coupled type IIB models, where D7 brane 
stacks and their intersections are responsible for non-Abelian gauge symmetries 
and charged matter. We approach such compactifications from the F-theory perspective
\cite{v96,s97}, where the value of the type IIB dilaton is encoded in the complex structure 
of a torus attached to every point of the type IIB manifold. D7 branes are 
characterised by the degeneration loci of this torus fibration and 
non-Abelian gauge symmetries arise if the degeneration is so bad that 
the 8d compact space develops a singularity. Such singularities are 
associated with the shrinking of M-theory cycles, which is easy to 
ensure by the flux choice. In addition, in weakly coupled situations, the 
periods of  M-theory cycles measure the relative positions of D7 branes. 

Working in the simple setting of F-theory on $K3\times K3$ (which 
corresponds to type IIB on $K3\times T^2/\mathbb{Z}_2$), we are able to demonstrate 
how fluxes stabilise D7 branes or stacks of D7 branes in a completely explicit
fashion. This allows us to select a specific desired gauge group by the choice 
of flux numbers. As explained above, this is the same procedure required for 
the flux stabilisation of non-Abelian gauge symmetries in the 
(non-perturbative) F-theory context, which has recently attracted significant 
attention in the context of GUT model building \cite{bhv08,dw08,dwbis08,hv08,mss08}. We therefore
expect that straightforward generalisations of our methods will be useful both for more
complicated D7 brane models as well as for their non-perturbative F-theory cousins. 

Moduli stabilisation by fluxes in M-theory on $K3\times K3$ has been studied extensively in the
past, especially in relation with the type IIB dual 
(see, e.g.~\cite{gkt04,lmr05,drs99}). In our work we derive the flux potential for the geometric
moduli from dimensional reduction. We express it in a form manifestly invariant under the $SO(3)$
symmetry of the $K3$ moduli space. In this form, it is immediate to see how the minimisation condition
is translated into a condition on fluxes and on geometric data of the two $K3$'s. We find all
Minkowski minima, both supersymmetric and non-supersymmetric. 
An analogous explicit search for (supersymmetric) flux vacua has been reported in~\cite{ak05}. Our
results are more  general since we do not restrict  ourselves to
attractive $K3$ surfaces, where a maximal number of integral 2-cycles are holomorphic.\footnote{At  
a technical level, this means that only a discrete set of values are allowed
for the various complex structure moduli. There is then also only a 
very restricted set of fluxes which are suitable for stabilising such points.}

Our analysis of moduli stabilisation is also more explicit than the previous works on $K3\times K3$,
since we use a parameterisation of D7-brane
motion by the size of integral two-cycles, as derived in~\cite{bht08}. Thus, at least
in the weak coupling limit, we have a simple geometric interpretation for 
every integral basis cycle. Using the choice of flux numbers, this gives us 
full control over the positions of 4 O7 planes and 16 D7 branes moving on a 
$\mathbb{CP}^1$ base (corresponding to type IIB on $T^2/\mathbb{Z}_2$). % as well as the complex 
%structure of the second `non-F-theory' $K3$. 

Our techniques can be used to study the stabilisation of all the gauge groups that can be realised
by F-theory on $K3$. It turns out that tadpole cancellation is very restrictive and allows only very
special flux choices. 

We begin our analysis in Section~2 with a derivation of the $K3\times K3$ flux
potential, which closely follows the generic Calabi-Yau derivation 
of~\cite{gvw99, bb96, hl01}. We emphasise the fact that, 
due to the hyper-K\"ahler structure of $K3$, its geometric moduli space can 
be visualised by the motion of a three-plane in the 22-dimensional space of 
homology classes of two-cycles. This three-plane is spanned by the real and 
imaginary parts of the holomorphic 2-form and by the K\"ahler form. The
resulting $SO(3)$ symmetry of the geometric moduli is manifest in the
expression for the scalar potential we arrive at. The three-dimensional theory also has a number of
gauge fields, and the flux induces mass terms for some of them, which we derive explicitly. This
breaking of gauge symmetries can be understood in the dual type~IIB picture as the gauging of some
shift symmetry in the flux background.

In Section~3 we analyse the minima of the above flux potential. 
To preserve four-dimensional Poincar\'e invariance, we consider 4-form 
fluxes that belong to $H^2(K3)\otimes H^2(K3)$.
A flux of this form gives rise to a linear map between the spaces of two-cycles of
the two $K3$'s: integrating the flux on a 2-cycle of one $K3$ we get a 2-form on the other 
$K3$ (which is Poincar\'e dual to a 2-cycle). Minkowski vacua arise if the flux 
maps the three-planes determining the metric of the $K3$'s onto each other. 
We derive the conditions the flux matrix has to satisfy in order 
for two such planes to exist and to be completely fixed by the choice of 
fluxes. Furthermore, we clarify the more restrictive conditions under which 
the plane determined by the flux is consistent with the F-theory limit. In 
this case, the plane cannot be fixed completely. The unfixed moduli 
correspond to Wilson lines around the $S^1$ of the type IIB model which 
decompactifies in the F-theory limit. These degrees of freedom are not part 
of the moduli space of type~IIB compactified to four dimensions, as they 
characterise the (unphysical) constant background value of one component of 
the four-dimensional vector fields. In fact, the corresponding propagating 
degrees of freedom become part of the four-dimensional vector fields 
(see~\cite{v08} for a comprehensive analysis of the duality map between the 
4d fields of M-theory on $K3\times K3$ and type IIB string theory on 
$K3\times T^2/\mathbb{Z}_2$). 

The main point of our paper, the explicit stabilisation of D-brane positions, 
is the subject of Section~4. After recalling the parameterisation of D7-brane 
motion in terms of M-theory cycles derived in~\cite{bht08}, we provide 
explicit examples of flux matrices which fix situations with gauge symmetries 
$SO(8)^4$, $SO(8)^3\times SO(6)$ and $SO(8)^3\times SO(4)\times SU(2)$. In all cases we also fix 
the complex structure moduli of the lower $K3$. The first case corresponds to the orientifold, where
4 D7 branes lie on top of each O7 plane. In the second case, one  D7~brane is moved away from an
O~plane. Finally, in the third case, a stack of two D7~branes is separated from one of the O~planes.
In these examples almost all the K\"ahler moduli (which correspond to deformations of the lower $K3$
and do not  affect the positions of the D7 branes) are not stabilised. When one of them is
stabilised, a K\"ahler modulus of the upper $K3$ is stabilised, too. As mentioned before, this
corresponds to some gauge field becoming massive. To clarify this point, we 
present two examples where one of the D7 branes is fixed at a certain 
distance from its O~plane: In the first example, one further K\"ahler modulus 
is fixed, breaking the $U(1)$ gauge group. This phenomenon of gauging by 
fluxes is common in flux compactifications\cite{aaf03,aaft03,hkl06,jl04}. In the
second example, we stabilise the single D7 brane without fixing further 
K\"ahler moduli and hence without gauge symmetry breaking. We also provide 
an example where almost all moduli are fixed. In this case, only the fibre  
volume of F-theory, the volume moduli of the two $K3$s, and three metric 
moduli of the lower $K3$ remain undetermined.

Section~5 contains a brief discussion of supersymmetry. Generically, we obtain 
$\Nn=0$ vacua of no-scale type. For specific, non-generic choices of the flux 
matrix, we find three-dimensional ${\cal N}=2$ or ${\cal N}=4$ supersymmetry.
To determine the amount of surviving supersymmetry, it suffices to know the 
eigenvalues of the flux matrix restricted to the two three-planes.

After summarizing our main results in Section~6, we collect some technical 
issues in the appendices. Appendix~\ref{sec:K3} contains some basic definitions 
concerning the geometry of $K3$, Appendix~\ref{VWWhat} gives the flux potential in terms 
of the two superpotentials of M-theory compactifications. Some facts about self-adjoint
operators on spaces with indefinite metric are collected in Appendix~\ref{app:linalg}. 
Finally, Appendix~\ref{ftheorypoint} supplies some further details concerning the F-theory limit,
especially the way in which certain M-theory moduli are lost in the F-theory 
limit.

\section{\boldmath$K3$ Flux Potential\label{sec:fluxpotential}}

In this chapter we compactify M-theory to three dimensions on $K3\times K3$ 
and analyse the effects of four-form flux. The main new points of our 
presentation are the following: We maintain a manifest $SO(3)\times SO(3)$ 
symmetry of the moduli space of $K3\times K3$ in the calculation of the 
potential in Section~\ref{sec:potential}. Furthermore, we explicitly derive 
the flux-induced masses for the vector fields arising from the three-form 
$C_3$ in Section~\ref{sec:vectormass}.

\subsection[M-Theory on $K3\times K3$]{M-Theory on \boldmath$K3\times K3$}
The compactification of M-theory on a generic four-fold is described in 
detail in\cite{bb96}. Here we specialise to the case of $K3\times K3$. To 
distinguish the two $K3$'s, we write the compactification manifold as 
$K3\times \Kt$. Correspondingly, all quantities related to the second $K3$ 
will have a tilde.

The relevant M-theory bosonic action is\cite{cjs78}
\begin{align}\label{MthAction}
  \begin{split}
    S_\text{M} &= \frac{2\pi}{\ell_M^9} \left\{ \int \d^{11}x \sqrt{-g} 
    \left( R -\frac{1}{2} |F_4|^2
    \right)- \frac{1}{6} \int  C_3\wedge F_4\wedge F_4 \right\}\\
    &\quad +\left(\frac{2\pi}{\ell_M^3}\right) \left(\int 
    C_3\wedge I_8(R) + \int
      \d^{11} x \sqrt{-g}\, J_8\!\left(R\right)\right)\,,
  \end{split}
\end{align}
where $\ell_M$ is the eleven-dimensional Planck length, $F_4=dC_3$, and 
$I_8(R)$ and $J_8\!\left(R\right)$ are polynomials of degree 4 in the 
curvature tensor \cite{dlm95,t00}. When we compactify on $K3\times \Kt$, we obtain a 
three-dimensional theory with eight supercharges, i.e.\ $\Nn=4$ in three 
dimensions. This can be inferred from the fact that each $K3$ has holonomy
group $SU(2)$ and correspondingly two invariant spinors. 

Let us analyse the geometric moduli. $K3$ is a hyper-K\"ahler manifold: its 
metric is defined by three two-forms $\omega_i$ in $H^2(K3)$ plus the overall 
scale. $H^2\!\left(K3\right)$ is a 22-dimensional vector space equipped with 
a natural scalar product,\footnote{
Throughout
this work, we freely identify forms, their cohomology classes, the 
Poincar{\'e}-dual cycles and their homology classes.
}
\begin{align}\label{K3modmetric}
  v \cdot w \equiv \int_{K3} v \wedge w \qquad \forall\; 
  v,w\in H^2\!\left(K3\right)\,,
\end{align}
which has signature $(3,19)$, i.e.\ there are three positive-norm directions. 
The three vectors $\omega_i$ defining the metric must have positive norm and 
be orthogonal to each other. Hence they can be normalised according to 
$\omega_i \cdot \omega_j = \delta_{ij}$. The K\"ahler form and 
holomorphic two-form and can then be given as
\begin{align}
  j&= \sqrt{2 \nu}\,\omega_3\,, & \omega&=\omega_1 +\I \omega_2\,.
\end{align}
This definition is not unique: we have an $S^2$ of possible 
complex structures and associated K\"ahler forms. Each of them defines the 
same metric, which is then invariant under the $SO(3)$ that rotates the 
$\omega_i$'s.

The motion in moduli space can now be visualised as the motion of the 
three-plane $\Sigma$ spanned by the $\omega_i$'s, which is characterised 
by the deformations of the $\omega_i$ preserving orthonormality. The 
corresponding $\delta \omega_i$ are  in the subspace orthogonal to 
$\Sigma$, which is 19-dimensional. Together with the volume, this
gives $3\cdot 19+1=58$ scalars in the moduli space of one $K3$. The same 
parameterisation can be used for the second $K3$, where the corresponding 
scalars are $\nut$ and the components of $\delta\omegat_j$. Altogether one 
finds $58+58=116$ scalars from the metric on $K3\times \Kt$. Furthermore, 
since $K3$ has no harmonic one-forms, there are no 3d vectors coming from 
the metric.

\subsection{The Scalar Potential\label{sec:potential}}

We now allow for an expectation value for the field strength $F_4$
of the form 
\begin{align} \label{4formflux}
  \left<F_4\right> \equiv G_4= G^{I\Lambda} \eta_I\wedge\etat_\Lambda\,,
\end{align}
where $\{\eta_I,\etat_\Lambda\}$ (with $I,\Lambda=1,...,22$) is an integral 
basis of $H^2(K3)\times H^2(\Kt)$. The flux satisfies a (Dirac) 
quantisation condition\footnote{
The 
precise quantisation condition for a generic fourfold $Y$ is $\ell_M^{-3}\,
[G_4]-\frac{p_1}{4} \in H^4(Y,\mathbb{Z})$, where $p_1$ is the first 
Pontryagin class\cite{w96}. Since $\frac{p_1}{2}$ is even for $Y=K3\times K3$, the 
quantisation condition becomes simply $\ell_M^{-3}\,[G_4] \in 
H^4(Y,\mathbb{Z})$.
}:
$\ell_M^{-3} G^{I\Lambda}\in\mathbb{Z}$. In the following, we will always 
denote this type of flux by $G_4$ while the generic four-form field strength 
will be $F_4$. 

The flux potential for the moduli is found by reducing the M-theory action.
In the presence of fluxes, the solution to the equations of motion is a 
warped product of a Calabi--Yau fourfold and a three-dimensional non-compact 
space\cite{bb96,drs99,gkp01}. In the following, we neglect backreaction and
work with the undeformed Calabi--Yau space $K3\times \Kt$ as the internal 
manifold. The underlying assumption is that, in analogy to\cite{gkp01}, 
for any zero-energy minimum of the unwarped potential a corresponding 
zero-energy warped solution will always exist. After Weyl rescaling, the 
potential is given by\cite{hl01}
\begin{align}\label{ModPotential}
  V= \frac{4\pi}{\ell_M^9}\frac{1}{4\Vv^3}\left(\,\int_{\mathrlap{
  \mspace{20mu}K3\times\Kt}}\d^8\xi
  \sqrt{g^{(8)}} |G_4|^2 - \frac{\ell_M^6}{12}\, \chi\right) \,,
\end{align}
where $\chi$ is the Euler number of the compact manifold. For
$K3\times\Kt$, it is $\chi=24^2=576$. Given our previous discussion of $K3$ 
moduli space, we expect that \eqref{ModPotential} will be invariant under 
$SO(3)\times SO(3)$ rotations once we express the metric in terms of 
$\omega_i$ and $\omegat_j$. 

In the absence of spacetime-filling $M2$ branes, the cancellation of 
$M2$-brane-charge on the compact manifold $K3\times \Kt$ requires\cite{bb96}
\begin{align}\label{TadpoleCanc}
 \frac{1}{2\,\ell_M^6}\int G_4\wedge G_4 =  \,\frac{\chi}{24} \:.
\end{align}
This allows us to express the second term in~\eqref{ModPotential} through 
the flux. It is convenient to set $\ell_M=1$ and to introduce a 
volume-independent potential $V_0$ by writing $V=\frac{2\pi}{\Vv^3} \, V_0$. 
Here $\Vv=\nu\nut$ is the volume of $K3\times\Kt$. Our result now reads 
\begin{equation}\label{V0expr}
 V_0 = \frac12 \int_{\mathclap{K3\times\Kt}} (G_4\wedge \ast G_4 - 
G_4\wedge G_4)\,,
\end{equation}
with $G_4$ given by \eqref{4formflux}. 

On $K3$, each $\eta_I$ can be split into a sum of two vectors, parallel and 
perpendicular to the 3-plane $\Sigma$:
\begin{align}\label{SplitEtaI}
  \eta_I = \sum_i(\eta_I\cdot \omega_i) \,\omega_i + \Pp[\eta_I]=
  \eta_I^\parallel+\eta_I^\perp \:.
\end{align}
Here $\Pp$ is the projector on the subspace orthogonal to $\Sigma$. The first
term, which corresponds to the projection on $\Sigma$, has been given in a 
more explicit form using the orthonormal basis $\omega_i$ of the $\Sigma$ 
plane for later convenience. The two terms of~\eqref{SplitEtaI} represent a 
selfdual and an anti-selfdual two-form\cite{gvw99}, allowing us to write the 
Hodge dual of a basis vector as
\begin{align}\label{HodgeonSplit}
  \ast_{\scriptscriptstyle K3}\,\eta_I =  \eta_I^\parallel-\eta_I^\perp \,. 
\end{align}
The same applies to $\widetilde{K3}$. 

If we insert \eqref{4formflux} and \eqref{SplitEtaI} into the 
expression~\eqref{V0expr} for $V_0$ and we use the relation  
\eqref{HodgeonSplit} for the action of the Hodge~$\ast$, we find
\begin{align}
  \begin{split}
    V_0 &= - \left\{\eta_I^\parallel\cdot \eta_J^\parallel
      \left(\left(G^{I\Lambda}\etat_\Lambda\right)^\perp \!\cdot
      \left(G^{J\Sigma}\etat_\Sigma\right)^\perp\right)\right.\\
  &\quad \mspace{35mu} \left.+ \left(\left(\eta_IG^{I\Lambda}\right)^\perp 
      \!\cdot\left(\eta_J G^{J\Sigma}\right)^\perp\right) 
  \etat_\Lambda^\parallel\cdot
        \etat_\Sigma^\parallel \right\} \,.
  \end{split}
\end{align}
Since 
\begin{align}
  \eta_I^\parallel\cdot \eta_J^\parallel&= \sum_i \left(\eta_I\cdot 
  \omega_i\right) \left(\eta_J\cdot \omega_i\right) \,,
\end{align}
we can write $V_0$ as
\begin{align}
  \begin{split}
    V_0 &=-\left\{ \sum_{i} \Ppt[G^{I\Lambda}\,(\eta_I\cdot\omega_i)\,
    \etat_\Lambda] \cdot \Ppt[
      G^{J\Sigma}\,(\eta_J\cdot\omega_i)\,\etat_\Sigma] \right.\\ 
    & \left.\mspace{153mu}+ \sum_{j}\Pp[G^{I\Lambda}\,(\etat_\Lambda\cdot
      \omegat_j)\,\eta_I] \cdot \Pp[
      G^{J\Sigma}\,(\eta_\Sigma\cdot\omegat_j) \,\eta_J] \right\} \,.
  \end{split}
\end{align}

To write it in a more compact form, we define two natural homomorphisms 
$G:H^2(\Kt) \rightarrow H^2(K3)$ and $G^a:H^2(K3)\rightarrow H^2(\Kt)$ by
\begin{align}
  G\, \tilde{v} &=\int_\Kt G_4\wedge \tilde{v} = (G^{I\Lambda} 
  \Mt_{\Lambda\Sigma} \tilde{v}^\Sigma)\, \eta_I\,, & G^a v &=\int_{K3} 
  G_4\wedge v= (v^J M_{JI} G^{I\Lambda}) \etat_\Lambda \,.
\end{align}
where $v=v^J\eta_J\,\in H^2(K3)$, $\tilde{v}=\tilde{v}^\Sigma\etat_\Sigma\, 
\in H^2(\Kt)$ and $M_{IJ}$, $\Mt_{\Lambda\Sigma}$ represent the metrics in 
the bases $\eta_I$, $\etat_\Lambda$. The operator $G^a$ is the adjoint 
of $G$, i.e. $(v\cdot G \tilde{v})=(G^a v,\tilde{v})$. The matrix 
components of these operators are $G^I_{\phantom{I}\Sigma}\equiv G^{I\Lambda} 
\Mt_{\Lambda\Sigma}$  and $\left(G^a\right)^\Sigma_{\phantom{\Sigma}I}
\equiv \left(G^T\right)^{\Sigma J}M_{JI}$. 

The moduli potential is then given by 
\begin{align}\label{potentialV}
  V = -\frac{2\pi}{\Vv^3} \left( \sum_i \left\| \Ppt[G^a\omega_i] \right\|^2 
  + \sum_j \left\| \Pp[G\,\omegat_j] \vphantom{\Ppt}\right\|^2\right) \,.
\end{align}
As expected, it is symmetric under $SO(3)$ rotation of the $\omega_i$'s and 
of the $\omegat_i$'s\footnote{
The 
projectors $\Pp$ and $\Ppt$ are obviously symmetric as they project onto the 
space orthogonal to all the $\omega_i$'s.}.

This potential is positive definite since the metrics for $H^2(\Kt)$ and 
$H^2(K3)$ defined in \eqref{K3modmetric} are negative definite on the subspace 
orthogonal to the $\omega_i$'s and  the $\omegat_i$'s. We note also that the 
volumes of the two $K3$'s are flat directions parameterizing the degeneracy 
of the absolute minimum of the potential, in which $V=0$.

We can also rewrite this potential expressing the projectors through the $\omega$'s:
\begin{align}\label{potentialV2}
  V = \frac{2\pi}{(\nu\nut)^3} \left( -\sum_i \left\| G^a\omega_i \right\|^2 
  - \sum_j \left\| G\,\omegat_j \right\|^2  + 2 \sum_{i,j}(\omegat_j\cdot 
    G^a \omega_i)(\omega_i \cdot G\, \omegat_j)\right) \,.
\end{align}
This is again manifestly symmetric under $SO(3)$ rotations. The potential 
can also be expressed in terms of two superpotentials (see 
Appendix~\ref{VWWhat}). We will not need this formulation in the following. 

In \eqref{4formflux} we have only considered fluxes $G_4$ with two legs on 
each  $K3$. More generally the flux could be of this form:
\begin{align}\label{4formfluxbis}
   \left<F_4\right>=G_4 + \mathcal{G} \rho + \widetilde{\mathcal{G}} \rhot\,,
\end{align}
where $\rho$ and $\rhot$ are the volume forms on $K3$ and $\Kt$.\footnote{
The 
normalisation is $\int_{K3} \rho   =\int_{\Kt} \rhot  =1$.
}
To obtain the general potential, we need to compute $\ast \left<F_4\right>$. 
Using our previous result for $\ast G_4$ and the Hodge duals 
\begin{align}
  \ast\rho = \frac{\nut}{\nu} \,\rhot \qquad\mbox{and}\qquad \ast \rhot = 
  \frac{\nu}{\nut} \, \rho \,
\end{align}
of $\rho$ and $\rhot$, we find 
\begin{align}
  \begin{split}
    V_\text{new} &= \frac{\pi}{(\nu\nut)^3}\int_{\mathclap{K3\times\Kt}} 
   \left(F_4\wedge \ast F_4 - F_4\wedge F_4 \right)  \\
    &=  \frac{2\pi}{(\nu\nut)^3}\left\{V_0 + \frac{1}{2} \mathcal{G}^2 
    \left(\frac{\nut}{\nu}\right)  +\frac{1}{2} \widetilde{\mathcal{G}}^2 
    \left(\frac{\nu}{\nut}\right) - \mathcal{G} \widetilde{\mathcal{G}} 
    \right\}\,.
  \end{split}
\end{align}
With the substitutions $\mathcal{V}=\nu\nut$ and $\xi=\sqrt{\frac{\nut}
{\nu}}$, the potential can be concisely written as
\begin{align}
  V_\text{new} = \frac{2\pi}{\mathcal{V}^3} \left\{V_0+ \frac{1}{2} 
  \left(\mathcal{G} \,\xi -\widetilde{\mathcal{G}}\, \frac{1}{\xi}\right)^2 
  \right\} \,.
\end{align}
This potential is still positive definite and has minima at points where it 
vanishes, but it now has only one unavoidable flat direction, the overall 
volume of $K3\times \Kt$. The ratio of the volumes is fixed at 
$\xi^2=\widetilde{\mathcal{G}}/\mathcal{G}$.

\subsection{Gauge Symmetry Breaking by Flux\label{sec:vectormass}}
In our context, F-theory emerges from the duality between 
M-theory on $K3\times\Kt$, with $\Kt$ being elliptically fibred, and 
type IIB on \mbox{$K3\times T^2/\mathbb{Z}_2\times S^1$}. The F-theory limit 
consists in taking the fibre volume to zero on the M-theory side, and in 
taking the radius of the $S^1$ to infinity on the type IIB side (see 
Section~\ref{sec:ftheory} for the details of this limit). Before analysing 
the effect of gauge symmetry breaking by fluxes, we recall the different 
origins of four-dimensional gauge fields in type IIB and in F-theory. 

Type~IIB theory on $K3\times T^2/\mathbb{Z}_2$ contains 16 vectors from 
gauge theories living on D7 branes and 4 vectors from the reduction of 
$B_2$ and $C_2$ along one-cycles of $T^2/\mathbb{Z}_2$. In three 
dimensions, one then has 20 three-dimensional gauge fields and 20 scalars 
corresponding to Wilson lines along the $S^1$. In the F-theory limit, these 
scalars combine with the vectors to give the required 20 four-dimensional 
vector fields. 

In M-theory on $K3\times\Kt$, vectors arise from the reduction of the 
three-form $C_3$ along two-cycles in $K3$ or $\Kt$. Since we are in three 
dimensions we have the freedom to dualise some of these vectors, treating 
them as three-dimensional scalars. To match the type~IIB description, the 
correct choice is to treat only the fields coming from the reduction of 
$C_3$ on two-cycles of $\Kt$ as vectors\footnote{A 
simple intuitive argument for this choice can be given by comparing the 
seven-dimensional theories coming from M-theory on $\Kt$ and type IIB on 
$T^2/\mathbb{Z}_2\times S^1$. In M-theory, we have seven-dimensional vectors 
associated with two-cycles stretched between the pairs of degeneration loci 
of the fibre (which characterise D7 branes). In type IIB, the corresponding 
vectors come directly from the D7-brane world-volume theories. The fact that 
they are associated with branes rather than with pairs of branes is simply a 
matter of basis choice in the space of $U(1)$s.}.
This reduction gives 22 vectors in three dimensions. However, since $\Kt$ is 
elliptically fibred, there are two distinguished two-cycles: the base and the 
fibre. They require a special treatment in the F-theory limit and, as a 
result, three-dimensional vectors arising from these two cycles do not become 
four-dimensional gauge fields in the F-theory limit. Instead, one of them 
corresponds to the type~IIB metric with one leg on the $S^1$, while the 
other is related to $C_4$ with three legs on $T^2/\mathbb{Z}_2\times 
S^1$\cite{v08}. We will not consider these two vectors in the following and 
focus on the remaining 20 three-dimensional vectors associated with the 
reduction of $C_3$ on generic two-cycles of $\Kt$. 

Each of these vectors absorbs a three-dimensional scalar (corresponding to a 
Wilson line degree of freedom on the type IIB side) to become a 
four-dimensional vector. These 20 scalars come from the metric moduli space 
of $\Kt$. More precisely, 18 arise from the variations $\delta\omegat_3$ of 
the K\"ahler form in directions orthogonal to the three-plane and to the 
base-fibre subspace\footnote{
For 
an elliptically fibred $\Kt$, two directions of the three-plane are orthogonal 
to base and fibre subspace, while $\omegat_3$ has a component along the 
base-fibre subspace. This explains the above number of independent 
variations as $18=22-(3+2-1)$. The variation of $\omegat_3$ within the 
base-fibre subspace corresponds to part of the metric in the F-theory limit. 
}. 
The two remaining scalars come from variations $\delta\omegat_1$ and 
$\delta\omegat_2$ of the holomorphic two-form which lie in the base-fibre 
subspace and are orthogonal to $\omegat_3$. For a detailed analysis of the 
matching of fields on both sides of the duality, see~\cite{v08}.

Given these preliminaries, it is now intuitively clear why F-theory flux 
generically breaks gauge symmetries: The flux induces a potential for the 
metric moduli, making them massive. This applies, in particular, to those 
moduli which become vector degrees of freedom in four dimensions. Hence, 
the full four-dimensional vector becomes massive by Lorentz 
invariance\footnote{
Correspondingly 
in type IIB,  putting two-form flux on certain cycles of wrapped D7 branes 
breaks the gauge symmetry of the brane\cite{hkl06,aaf03,aaft03,jl04}.}.

To derive the vector mass term explicitly, we begin by writing $C_3$ in 
the form
\begin{align}\label{eq:C3KK}
  C_3&=C_1^I\wedge \eta_I + \Ct_1^\Sigma\wedge\etat_\Sigma+C_3^\text{flux}\,. 
\end{align}
Here $\eta_I$ and $\etat_\Sigma$ are basis two-forms on the $K3$ factors, 
$C_1^I$ and $\Ct_1^\Sigma$ are one-form fields in three dimensions, and 
$C_3^\text{flux}$ is the contribution responsible for the four-form flux 
(which is only locally defined). As before, the flux is given by $G_4=
G^{I\Sigma}\eta_I\wedge\etat_\Sigma=\d C_3^\text{flux}$. 

In the reduction of the action, the $\int \left|F_4\right|^2$ term leads to 
the flux term $\int \left|G_4\right|^2$ (which is irrelevant for our present 
discussion) and to kinetic terms for $C_1^I$ and $\Ct_1^\Sigma$. The metric 
for the kinetic terms is given by
\begin{align}
  &\int _{\mathclap{K3\times\Kt}} \eta_I\wedge*_8\eta_J=\nut \int_{K3}
  \left(\eta_I^\parallel \wedge \eta_J^\parallel -\eta_I^\perp\wedge 
  \eta_J^\perp \right)=\nut \, g_{IJ} \,,\\ 
  &\int_{\mathclap{K3\times\Kt}} \etat_\Lambda\wedge*_8\etat_\Sigma =\nu
  \int_{\Kt}\left(\etat_\Lambda^\parallel \wedge \etat_\Sigma^\parallel -
  \etat_\Lambda^\perp\wedge \etat_\Sigma^\perp \right)=\nu\, 
  \gt_{\Lambda\Sigma}\,. 
\end{align}
We have split off the volume dependence, so that $g_{IJ}$ and 
$\gt_{\Lambda\Sigma}$ are dimensionless. Note that these metrics are positive 
definite since the subspace orthogonal to the three-plane has 
negative-definite metric. Note also that there is no kinetic mixing between 
the $C_1^I$ and the $\Ct_1^\Sigma$ since $\int \eta_I\wedge*_8\etat_\Sigma=0$.

We now turn to the Chern--Simons term $\int C_3\wedge F_4\wedge 
F_4$. Evaluating this term with $C_3$ of the form~(\ref{eq:C3KK}), we see that 
the contribution $\int C_3^\text{flux} \wedge F_4\wedge F_4$ vanishes: 
$C_3^\text{flux}$ has three legs on $K3\times\Kt$, so $F_4\wedge F_4$ would
need to have three legs on $\mathbb{R}^{1,2}$ and five legs on $K3\times\Kt$. 
This is, however, inconsistent with~(\ref{eq:C3KK}). The other contributions 
give
\begin{align}
  \int  C_3\wedge F_4 \wedge F_4 &= \int_{\mathbb{R}^{1,2}}2 G_{I\Sigma} 
  \left(C_1^I\d \Ct_1^\Sigma +\Ct_1^\Sigma \d C_1^I  \right) \,.
\end{align}
Thus, the flux matrix $G_{I\Sigma}= M_{IJ}G^{J\Lambda} \Mt_{\Lambda\Sigma}$ 
couples $C_1^I$ and $\Ct_1^\Sigma$. (Note that flux proportional to the 
volume forms of $K3$ and $\Kt$ would, in addition, lead to couplings 
$\sim C_1^I\d C_1^J$ and $\sim \Ct_1^\Sigma\d \Ct_1^\Lambda$.) 

We have now arrived at the three-dimensional effective action
\begin{align}\label{eq:3dactionflux}
  \begin{split}
    S_C^{(3)}&= \int_{\mathbb{R}^{1,2}} \left\{\nut g_{IJ}\, \d C_1^I \wedge 
    *\d C_1^J +\nu \gt_{\Lambda\Sigma}\, \d
    \Ct_1^\Lambda \wedge *\d \Ct_1^\Sigma\vphantom{\frac{2}{3}}\right.\\
    &\quad\mspace{40mu}\left.+\frac{2}{3}G_{I\Lambda} \left(C_1^I\wedge 
    \d \Ct_1^\Lambda+\Ct_1^\Lambda \wedge \d  C_1^I\right)\right\} \,. 
  \end{split}
\end{align}
As explained before, only the vectors $\Ct_1^\Sigma$ become four-dimensional 
vectors in the F-theory limit\cite{v08}. It is convenient to dualise the 
remaining vectors $C_1^I$, replacing them by scalars $C_0^I$. To this 
end, we turn the equation of motion, 
\begin{align}
  \d \left(* \d C_1^I + \frac{2}{3} \frac{1}{\nut} g^{IJ} G_{J\Sigma} \d 
  \Ct_1^\Sigma\right)\,=\,0\,,
\end{align}
into a Bianchi identity by defining $C_0^I$ through
\begin{align}
  * \d C_1^I+\frac{2}{3}\frac{1}{\nut}  g^{IJ} G_{J\Sigma} \Ct_1^\Sigma\,=\, 
  \d C_0^I \,.
\end{align}
It follows that the $C_0^I$ have to transform non-trivially under the gauge 
transformations of the $\Ct_1^\Sigma$:
\begin{align}\label{eq:dualgauge}
  \Ct_1^\Sigma&\longrightarrow  \Ct_1^\Sigma +\d \Lambda_0^\Sigma\,, & 
  C_0^I &\longrightarrow C_0^I +\frac{2}{3}\frac{1}{\nut} g^{IJ}G_{J\Sigma} 
  \Lambda^\Sigma_0\,.
\end{align}
In other words, the vectors $\Ct_1^\Sigma$ gauge shift symmetries of the 
scalars $C_0^I$, with the charges determined by the flux. 

The equation of motion of $C_0^I$ follows formally from $\d \d C_1^I=0$, the 
Bianchi identity of $C_1^I$. Since $**=-1$ on $\mathbb{R}^{1,2}$, we find 
\begin{align}
0=\d\d C_1^I=-\d ** \d C_1^I&=\d * \left(\d C_0^I -\frac{2}{3}\frac{1}{\nut} 
g^{IJ}G_{J\Sigma} \Ct_1^\Sigma\right)\,.
\end{align}
We now want to find a gauge invariant action from which this equation of 
motion can be derived. Such an action is given by 
\begin{align}
  S_C^\text{dual}&= \int_{\mathbb{R}^{1,2}} \d^3 x \sqrt{-g_3}\left\{ \nu \left|\d \Ct_1^\Sigma\right|^2 
  + \nut \left|\d C_0^I - \frac{2}{3} \frac{1}{\nut} g^{IJ} G_{J\Sigma} 
  \Ct_1^\Sigma \right|^2\right\}  \,. 
\end{align}
The corresponding Einstein-Hilbert term has the usual volume prefactor and can 
be brought to canonical form by a Weyl rescaling of the three-dimensional metric. This gives 
the kinetic term of the vectors a prefactor $(\nu\nut)\nu$, which we can 
absorb in a redefinition of $\Ct_1^\Sigma$. The resulting mass matrix has 
the form
\begin{align}
  m^2_{\Sigma\Lambda}&\sim \frac{1}{\left(\nu\nut\right)^3} G_{I\Sigma} 
G_{J\Lambda} g^{IJ} \,,
\end{align}
which is positive semidefinite since $g^{IJ}$ is a positive definite metric. 
The number of gauge fields which become massive is determined by the rank of 
the flux matrix. Comparing with Eq.~(\ref{potentialV}), we see that the 
masses are of the same order as the masses of the flux stabilised geometric 
moduli. This confirms the intuitive idea put forward at the beginning of 
this section: The vectors $\Ct_1^\Sigma$ and some geometric moduli are 
combined in the F-theory limit to produce four-dimensional vectors. For this to work in 
the presence of fluxes, both the three-dimensional vectors and scalars need to have the 
same flux-induced masses.

\section{Moduli Stabilisation\label{sec:modstab}}

In this section we turn to the flux stabilisation of moduli. First, we will analyse under which
conditions the potential~(\ref{potentialV}) has minima at $V=0$, and whether there are flat
directions. Then we will see which restrictions we have to impose in order to map the M-theory
situation to F-theory, and discuss possible implications for gauge symmetry breaking.

Let us first comment on the flux components which are proportional to the volume forms. In what
follows, we do not consider these components, in other words, we set
$\mathcal{G}=\widetilde{\mathcal{G}}=0$. The reason is that we want to end up with a
Lorentz-invariant four-dimensional theory. By going through the M-theory/F-theory duality
explicitly, one can see that this requires that the flux needs to have exactly one
leg in the fibre torus and hence two legs along each $K3$. Thus, we can without loss of generality
use a flux in the form of Eq.~(\ref{4formflux}), and the associated potential~(\ref{potentialV}).

\subsection{Minkowski Minima}
Clearly, the potential~(\ref{potentialV}) cannot stabilise the volumes $\nu$ and $\nut$. They are
runaway directions in general, and flat directions exactly if the term in brackets vanishes. This
term is a sum of positive definite terms, so each of these must vanish if we want to
realise a minimum with vanishing energy. Since each term contains a projection onto the subspace
orthogonal to the three-planes spanned by the $\omega_i$'s and $\omegat_i$'s, the bracket clearly
vanishes if and only if the flux homomorphisms map the three-planes into each other, though not
necessarily bijectively:
\begin{align}\label{eq:planetoplane}
  G \,\Sigmat &\subset \Sigma\,, & G^a \Sigma \subset \Sigmat\,.
\end{align}
Note that what is required is not merely the existence of three-dimensional subspaces which are
mapped to each other, but that both subspaces are positive-norm. If the metrics were positive definite,
this condition would be trivial since any real matrix can be diagonalised by choosing appropriate
bases in $H^2\!\left(K3\right)$ and $H^2\!\left(\Kt\right)$.

We will now show that the conditions~\eqref{eq:planetoplane} are equivalent to the conditions that the
map $G^a G$ is diagonalisable and all its eigenvalues are real and non-negative%
\footnote{Note that $G^a G$ maps
$H^2\!\left(\Kt\right)$ onto itself, so it makes sense to speak of eigenvalues and eigenvectors.
Note also, however, that although $G^a G$ is a selfadjoint operator, this does not imply that its
eigenvalues are real since the metric is indefinite. We have collected some facts about linear
algebra on spaces with indefinite metric in Appendix~\ref{app:linalg}.}.

Let us assume that there exist two three-planes $\Sigma$ and $\Sigmat$ such that the relations \eqref{eq:planetoplane} hold.
The cohomology groups can be decomposed into orthogonal subspaces, $H^2\!\left(K3\right)=\Sigma\oplus R$ and
$H^2\!\left(\Kt\right)=\Sigmat\oplus \Rt$, such that the metric \eqref{K3modmetric} defined by the wedge product is positive
(negative) definite on $\Sigma$ and $\Sigmat$ ($R$ and $\Rt$). The
conditions~(\ref{eq:planetoplane}) imply that $G$ and $G^a$ are block-diagonal, i.e.~we also
have $G\Rt\subset R$ and $G^a R\subset \Rt$.
It is then obvious that the selfadjoint operator $G^aG$ obeys\footnote{Similarly $G\,G^a$ obeys $G\,G^a \Sigma \, \subset \Sigma$.}
\begin{equation}\label{eq:planetoplane2}
  G^a G \Sigmat \, \subset \Sigmat\,. %\qquad \mbox{ and similarly } \qquad G\,G^a \Sigma \, \subset \Sigma \:.
\end{equation}
As each block is selfadjoint relative to definite metrics, $G^a G$ is diagonalisable with real and non-negative eigenvalues.
% Furthermore, for any $\et\in\Sigmat$
% and $\vt\in\Rt$ we have 
% \begin{align}
%   (G^a G \et \cdot \et)&= (G\et \cdot G\et) \geq 0\,,\\
%   (G^a G \vt \cdot \vt)&= (G\vt \cdot G\vt) \leq 0\,,
% \end{align}
% so $G^a G$ is indeed positive semidefinite. The same reasoning applies to $G G^a$. 

We now show that the converse also holds. Assume that $G^a G$ is diagonalisable
with non-negative eigenvalues\footnote{In this case, because of the non-degeneracy of the inner product, there alway exists a basis of non-null eigenvectors.}. This defines a decomposition of $H^2(\Kt)$ in  $\Sigmat\oplus \Rt$,
where $\Sigmat$ is the three-dimensional subspace given by the eigenvectors with positive norm.
The fact that $G^aG$ maps $\Sigmat$ into itself implies that $G$ maps positive norm vectors into
positive norm vectors: Indeed, give $\et\in \Sigma$, 
\begin{align}
   (G\et \cdot G\et)=(G^a G\et \cdot \et) \geq 0\,,
\end{align}
If $G^aG|_{\Sigmat}$ is invertible (non-zero eigenvalues in $\Sigmat$), then we can define
$\Sigma$ as the image of $G|_{\Sigmat}$. The fact that $\Sigmat$ is invariant under $G^aG$
implies that the image of $G^a|_\Sigma$ is $\Sigmat$ and \eqref{eq:planetoplane} is proved.
The case in which $G^aG$ has non-trivial kernel does not present any complication. Since the kernel of $G^aG$
coincides with the kernel of $G$ \footnote{%One might worry that for an $\et$ in the kernel of $G^a G$, it
%might happen that $G\et\neq 0$ is a lightlike vector in $H^2\!\left(K3\right)$. However, s
Since $G$ and $G^a$ are adjoint to each other, there is an orthogonal decomposition $H^2\!\left(K3\right)=\im G
\oplus {\rm Ker}\, G^a$. Take $\et \in {\rm Ker}\, G^aG$; since $G \et$ is both in $\im G$ and in ${\rm Ker} \, G^a$, it is the zero vector, proving that $\et \in {\rm Ker}\, G$.},
the image of $G|_{\Sigmat}$ is no more three dimensional. One then defines $\Sigma$ as the image of $G|_{\Sigmat}$
plus the positive norm vectors in the kernel of $G^a$.

To summarise, the conditions~(\ref{eq:planetoplane}) are equivalent to the condition that $G^a G$ is
diagonalisable and all its eigenvalues are real and non-negative. In this case (see Appendix~\ref{app:linalg})
the matrices for $G^aG$, $G$ and $G^a$ take the form
\begin{equation}\label{GaGCanonicalForm}
 (G^aG)_d = \diag \!\left(a_1^2,a_2^2,a_3^2,b_1^2,\dotsc,b_{19}^2\right) \qquad G_d=G^a_d= 
	\diag \!\left(a_1,a_2,a_3,b_1,\dotsc,b_{19}\right) 
\end{equation}
in appropriate bases, where $a_i^2$ are the eigenvalues of $G^aG$ relative to positive norm eigenvectors, while $b_c^2$ are relative to negative norm eigenvectors.

\

Finally, we want to see whether there are flat directions. The potential
has a flat direction, if there are infinitesimally different positions of the three-planes 
$\Sigmat,\Sigma$ which give Minkowski minima. Given a flux (such that $G^aG$ is diagonalizable with positive eigenvalues), the minima correspond to $\Sigmat$ ($\Sigma$) generated by the positive norm eigenvectors of $G^aG$ ($G\,G^a$). If all the eigenvalues are different from each other, there can be only three positive norm eigenvectors, and the minimum is isolated. If a positive norm and a negative norm eigenvector have the same eigenvalue, e.g. $a_1=b_1$, then a flat direction arises: Any three-dimensional space spanned by $\vt_{a_2}$, $\vt_{a_3}$ and $\vt_{a_1}'=\vt_{a_1}+ \epsilon \,\ut_{b_1}$ ($\epsilon \ll 1$) will give a different $\Sigmat$ that still satisfies the conditions \eqref{eq:planetoplane}. It is easy to see that an analogous flat direction develops for $\Sigma$. Note that if some $a_i$ are degenerate then the rotation of the vectors does not move the three-planes.

This shows that flat directions of the potential are absent if and only if the sets of eigenvalues
$\left\{a_i^2\right\}$ and $\left\{b_a^2\right\}$ are pairwise distinct.

\subsection{F-Theory Limit\label{sec:ftheory}}
We are interested in stabilising points in the moduli space of $K3\times \Kt$ which can be mapped to
F-theory. This means we require that $\Kt$ is an
elliptic fibration over a base $\mathbb{C}\Pp^1$, and that the fibre volume vanishes.

The first requirement means that $\Kt$ needs to have two elements $\widetilde{B}$ (the
base) and $\widetilde{F}$ (the fibre) in the Picard group, i.e.~two integral (1,1)-cycles, whose
intersection matrix is
\begin{equation}\label{eq:blockBF}
\left(\begin{array}{cc}
	-2 & 1 \\ 1 & 0\\
\end{array}\right)\,.
\end{equation}
Note that by a change of basis from $\left(\widetilde{B},\widetilde{F}\right)$ to
$\left(\widetilde{B}+\widetilde{F},\widetilde{F}\right)$, this intersection
matrix is equivalent to one $U$ block in the general form~(\ref{eq:UUUE8E8}) of the metric in an
integral basis. 

As  (1,1)-cycles, $\widetilde{B}$ and $\widetilde{F}$ must be orthogonal to the holomorphic two-form. In
our case this means that the $\widetilde{\Sigma}$ plane has a two-dimensional subspace orthogonal orthogonal to
$\left<\widetilde{B},\widetilde{F}\right>$. This subspace is spanned by the real and imaginary part of 
the holomorphic two-form $\omegat=\omegat_1+\I\omegat_2$.
On the other hand, $\left<\widetilde{B},\widetilde{F}\right>$ contains the third positive-norm
direction, so $\omegat_3$ cannot be also orthogonal to 
$\left<\widetilde{B},\widetilde{F}\right>$. For the following discussion it is convenient to
consider directly the K\"ahler form $\jt$ instead of 
$\nut$ and $\omegat_3$ separately. The K\"ahler form can be parametrised as
\begin{align}\label{eq:jtelliptic}
  \jt&= b \widetilde{B} + f\widetilde{F} + c^a \ut_a \,,
\end{align}
where $\ut_a$ is an orthonormal basis (i.e.~$\ut_a\cdot \ut_b=-\delta_{ab}$) of the space orthogonal to
$\widetilde{F}$, $\widetilde{B}$ and $\omegat$. This is the most general form of $\jt$ for an
elliptically fibred $\Kt$.

\

Now we turn to the second requirement: the fibre must have vanishing volume. This is what is called
the {\it F-theory limit}. For the K\"{a}hler form~(\ref{eq:jtelliptic}), we find the volumes of the
fibre and the  base to be\footnote{
More generally the volume of a two-cycle $C_2$ is given by the projection on the three-plane $\Sigma$, multiplied by the $K3$ volume:
\begin{equation}
  \rho\left( C_2 \right) = \nu^{1/2} \sqrt{\sum_{i=1}^3 (\omega_i\cdot C_2)^2} = \nu^{1/2} \Vert C_2|_{\Sigma} \Vert \:.\nonumber
\end{equation}
}
\begin{align}
  \rho\!\left(\widetilde{F}\right)&=\jt\cdot\widetilde{F} = b\,, &
  \rho\!\left(\widetilde{B}\right)&=\jt\cdot\widetilde{B} = f -2b \,.  
\end{align}
Hence, the F-theory limit involves $b\to 0$, and in this limit, the base volume will be given by
$f$. On the other hand, the volume of the entire $\Kt$ is
\begin{align}
  \frac{1}{2}\, \jt\cdot\jt &= b\left(f-b\right) - \frac{1}{2} c^a c^a\,.
\end{align}
This volume is required to be positive, so we get a bound on the $c^a$,
\begin{align}\label{eq:caconstraint}
  \frac{1}{2}c^a c^a &< b\left(f-b\right)\,.
\end{align}
Thus, in the F-theory limit we have to take the $c^a$ to zero at least as fast as $\sqrt{b}$.
Once the limit is taken, the volume of $\Kt$ vanishes and the K\"ahler form is given by
\begin{align}\label{jFth}
 \jt = f \widetilde{F}\,,
\end{align}
regardless of the initial value of the $c^a$.
 Note that the constraint~(\ref{eq:caconstraint}) is consistent with the
intuitive picture of the fibre torus shrinking simultaneously in both directions: The $c^a$ measure
the volume of cycles which have one leg in the fibre and one in the base, %(and that are orthogonal
%to the complex structure $\omegat$)
so they shrink like the square root of the fibre volume $b$.

The K\"ahler moduli space is reduced in the F-theory limit: We lose not only the direction along
which we take the limit, but also all transverse directions except for the base volume $f$, which becomes
the single K\"ahler modulus of the torus orbifold. In the duality to type IIB on $K3\times
T^2/\mathbb{Z}_2\times S^1$, the $c^a$ parametrise Wilson lines of the gauge fields along the $S^1$
as long as the fibre volume is finite. In the F-theory limit, which corresponds to the radius of the
$S^1$ going to infinity, the Wilson lines disappear from the moduli space. The propapagating degrees of freedom related to them 
combine with the three-dimensional vectors from the three-form $C_3$ reduced along two-cycles of
$\Kt$ to form four-dimensional vectors (cf.~Section~\ref{sec:vectormass}, see also~\cite{v08}). 

From this perspective, we see that it is important not to fix the modulus controlling the size of
the fibre. In fact, if we leave it unfixed, we have a line in the M-theory moduli space
corresponding to this flat direction of the potential. Of this line, only the point at infinity
($b=0$) corresponds to F-theory. This limit is singular in the sense that the F-theory
point is not strictly speaking in the moduli space of $\Kt$, but on its boundary. 
As we show in Appendix~\ref{ftheorypoint}, this point
is at infinite distance from every other point in the moduli space of $\jt$, and it actually
corresponds to the decompactification limit in type IIB.

To see which fluxes are compatible with this limit, we first note that there must be no flux along
either $\widetilde{B}$ or $\widetilde{F}$ because Lorentz invariance of the four-dimensional theory
requires that the flux must have exactly one leg along the fibre. This means that in a basis of
$H^2\!\left(\Kt\right)$ consisting of $\widetilde{B}$, $\widetilde{F}$ and orthogonal forms, the
flux matrix must be of the form
\newfont{\krass}{cmr17 scaled 3500}
\begin{align}
  G^{I\Sigma} &=
  \begin{pmatrix}
    0 &0 &  \\
    \vdots & \vdots & \smash{\raisebox{-5.5ex}{\text{\krass *}}} \\
    0 &0 &
  \end{pmatrix}\,.
\end{align}
This leads to a $G^a G$  with two rows and columns of zeroes, 
\begin{align}\label{eq:GaGF}
  G^aG &= \begin{pmatrix} 0 &0 & \dots &0 \\0 &0 & \dots &0 \\ \vdots & \vdots
    &\mspace{-20mu}\mathrlap{\smash{\raisebox{-6.5ex}{\krass *}}}&\\ 0& 0&  &\\
    \end{pmatrix}\,,
\end{align}
hence the direction along which we take the F-theory limit is automatically flat.

To discuss the matrix form of $G$, it is convenient to choose an equivalent basis for $H^2(K3)$, i.e. a basis containing
$B$ and $F$ and 20 orthogonal vectors.
%For simplicity, we will in the following choose a basis in $H^2\!\left(K3\right)$ in which the first
%two vectors are essentially equivalent to a $U$ block of~(\ref{eq:UUUE8E8}), i.e.~they are
%orthogonal to the rest and the  metric on this subspace is of signature $(1,1)$.
We then restrict to fluxes of the type 
\begin{align}
  G^{I\Sigma}=\begin{pmatrix} \label{eq:Fthflux}
    \begin{matrix}\textstyle 0&0\\0&0\end{matrix} & \text{\Large 0}\\
        \text{\Large 0}\rule[14pt]{.5pt}{0pt}  & G^{I\Lambda}_\text{F-th}
    \end{pmatrix}\,,
\end{align}
although this is not the most general form. Here, $G^{I\Sigma}_\text{F-th}$ is a $20\times20$
matrix which we will also call $G^{I\Sigma}$ for simplicity. %Note that
%these fluxes are consistent with $\Kt$ to be elliptically fibred.

\section{Brane Localisation\label{sec:movebrain}}

One of our aims is to find a flux that fixes a given configuration of
branes. The results obtained so far allow us to do that: As we will review in the next subsection, the
positions of the D7 branes are encoded in the complex structure $\omegat=\omegat_1+\I \omegat_2$ of
$\widetilde{K3}$\cite{bht08}. This can be understood as follows: We can find certain cycles which
measure the distance between branes. A given brane configuration can thus be characterised by the
volumes of such cycles. Most relevant for the low-energy theory is the question whether there are
brane stacks (corresponding to gauge enhancement) which is signalled by the vanishing of
interbrane cycles. So choosing a given brane configuration determines a set
of integral cycles which are to shrink, i.e.~which should be orthogonal to the complex structure%
\footnote{These are cycles with one leg in the base and one in the fibre and which are
orthogonal to $\omegat_3$ once we take the F-theory limit \eqref{jFth}.}. We want to find what is the flux that
fixes such a complex structure.
%A flux that fixes this
% configuration thus needs to have a certain 
% block-diagonal structure in a basis which includes the shrunk cycles. 

The flux needs to satisfy some constraints: It must be integral and it must satisfy the
tadpole cancellation condition~(\ref{TadpoleCanc}). The first condition means that the entries of
the flux matrix $G^{I\Sigma}$ in a basis of integral cycles must be integers.
The tadpole cancellation condition translates into a condition on the trace of $G^a G$,
\begin{align}\label{TadpoleCancCond}
  \tr G^a G &= \tr G^T\!\! M G \Mt = 48\,.
\end{align}
Of course, we also require that the flux gives Minkowski minima, i.e.~$G^aG$ needs to have only
non-negative eigenvalues. These conditions turn out to be rather restrictive, and a scan of all
$20\times20$-matrices is computationally beyond our reach. Fortunately, the block-diagonal structure
alluded to above allows us to restrict to smaller submatrices of size $2\times 2$ or $3\times 3$,
where an exhaustive scan is feasible.

\subsection{D-Brane Positions and Complex Structure \label{sec:BraneCS}}

In the weak coupling limit, in which the F-theory background can be described by
perturbative type IIB theory, the complex structure deformations of the upper $K3$ have an 
interpretation in terms of the movement of D-branes and O-planes on $\mathbb{C}\Pp^1$ \cite{s97}.
From the perspective of the elliptically fibred $\widetilde{K3}$, D-branes and O-planes are 
points on the $\mathbb{C}\Pp^1$ base where the $T^2$ fibre degenerates. The positions of these
points are encoded in the complex structure of $\widetilde{K3}$: The 18 complex structure
deformations\footnote{
These 
are the deformations of $\omegat_1$ and $\omegat_2$ in the space orthogonal
to $\widetilde{F}$ and $\widetilde{B}$.
} 
specify the 16 D-brane positions, the complex structure of $T^2/\mathbb{Z}_2 \sim \mathbb{C}\Pp^1$, and the value of the complex
dilaton. The map between the two descriptions is worked out in detail in \cite{bht08}. In the following, we summarise the basic results.

When several D-branes coincide, the $\widetilde{K3}$ surface develops singularities which reflect
the corresponding gauge enhancement \cite{w95,sen96,a96,gz97,l99,a00ag04}.
These singularities can also be seen to arise when the volume of certain cycles shrinks to zero:
\begin{align}
\int_{\gamma_i}\omegat=\int_{\Kt}\gamma_i\wedge\omegat=\gamma_i\cdot\omegat \longrightarrow 0\,.
\end{align}
Note that these cycles have one leg on the base and and one leg on the fibre torus, so
$\gamma_i\cdot \jt=0$ \footnote{Since we are interested in the F-theory limit, we will only consider
  vacua corresponding to $\jt$ being in the block 
  $\left< \widetilde{F},\widetilde{B}\right>$.}. Hence their volume is given by $\sqrt{\nut} \left|
  \gamma_i\cdot\omegat\right|$. 
If the $\gamma_i$ are integral cycles (for the structure of integral cycles on $K3$ see
Appendix~\ref{sec:K3}) with self-intersection $-2$, their shrinking produces a singularity that
corresponds to a gauge enhancement.
Since these cycles can be thought of as measuring distances between branes, this is equivalent to
D-branes that are coinciding. The Cartan matrix that displays the gauge enhancement is
given by the intersection matrix of the shrinking $\gamma_i$.

\begin{figure}[ht!]
  \begin{center}
    \includegraphics[height=6cm,]{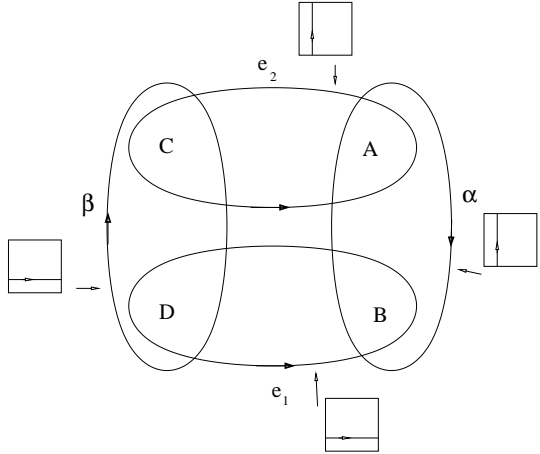}\caption{For an $SO(8)^4$ configuration,
      the two degrees of freedom of the O-planes and the dilaton are encoded in cycles   
      that surround two of the four blocks in the $\mathbb{C}\Pp^1$ base and wrap an arbitrary direction
      in the fibre torus. The four $SO(8)$ blocks are denoted by A,B,C and D. The four cycles displayed
      here form a basis that is dual to the four forms $e_i, \alpha$ and $\beta$. Note that we also have
      indicated the fibre part of each cycle. 
      \label{fourblocks}}
  \end{center}
\end{figure}

Let us consider an $SO(8)^4$ point at which $\widetilde{K3}$ degenerates to $T^4/\mathbb{Z}_2$. From the
D-brane perspective 
this corresponds to putting four D-branes on each of the four O-planes. In terms of the basis given 
in Appendix~\ref{sec:K3}, the complex structure of $\widetilde{K3}$ is given by
\begin{align}
\omegat_{SO(8)^4}=\frac{1}{2}\left(\alpha+Ue_2+S\beta-USe_1\right)\,.
\end{align}
For the sake of brevity we have introduced\footnote{Note that although $e_i$, $e^i$, $E_I$, $\alpha$
and $\beta$ are forms on $\Kt$, we omit the tildes to avoid unnecessary notational clutter.}
\begin{align}\label{eq:alphabeta}
  \alpha&\equiv2\left(e^1+e_1+W^1_IE_I\right)\,,&&& \beta &\equiv2(e^2+e_2+W^2_IE_I)\,,
\end{align}
where
\begin{align}
  W^1&=\left(0^4,\frac{1}{2}^4,0^4,\frac{1}{2}^4\right)\qquad \text{and} \qquad
  W^2=\left(1,0^7,1,0^7\right)
\end{align}
describe the mixing of cycles from the $U$ and $E_8$ blocks. Note that they can be interpreted as
Wilson lines, breaking $E_8\times E_8$ to $SO(8)^4$ in the duality to heterotic string theory. The
cycles that are dual to the forms  $e^i, \alpha$ and $\beta$ are shown in Figure~\ref{fourblocks}.
The parameter $U$ describes the positions of the four O-planes, which is equivalent to the
complex structure of the $T^2$ in type IIB before orientifolding. The dilaton, which is constant
in this configuration, is given by the complex structure of the fibre torus, $S$.

One can check that the cycles which have vanishing periods at an $SO(8)^4$ point span the lattice $D_4^4$.
They are given by
\begin{align}\label{AtoD}
  \text{
    \begin{tabular}[h]{c|c|c|c|c}
      & $A$&$B$&$C$&$D$\\
      \hline
      \rule[12pt]{0pt}{1pt}
      1&$E_{7}-E_{8}$&$-E_{15}+E_{16}$&
      $-e_2-E_{1}+E_{2}$&$e_2+E_{9}-E_{10}$\\
      2&$E_{6}-E_{7}$&$-E_{14}+E_{15}$ &
      $-E_{2}+E_3$&$E_{10}-E_{11}$\\
      3&$-e_1-E_{5}-E_{6}$&$e_1+E_{13}+E_{14}$&
      $-E_{3}+E_{4}$&$E_{11}-E_{12}$\\
      4&$E_{5}-E_{6}$&$-E_{13}+E_{14}$&
      $-E_{3}-E_{4}$&$E_{11}+E_{12}$.
    \end{tabular}}
\end{align}

\begin{figure}[ht!]
  \begin{center}
    \subfigure[]{\includegraphics[height=2cm]{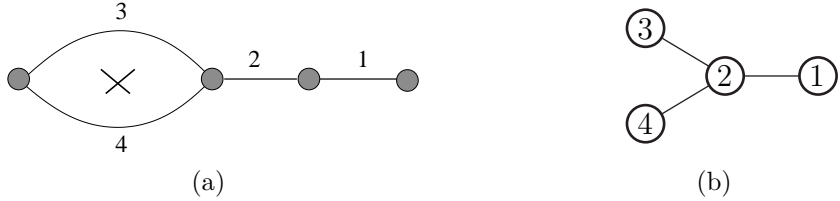}}
    \hspace{1cm}
    \subfigure[]{
    \begin{picture}(150,57)(0,0)
      %\EBox(0,0)(150,57)
      %\SetWidth{1}
      \Line(75,28.5)(110,28.5) \Line(75,28.5)(45,10.5) \Line(75,28.5)(45,46.5)
      \SetWidth{1}
      \BCirc(75,28.5){7} \BCirc(110,28.5){7} \BCirc(45,10.5){7} \BCirc(45,46.5){7}
      \Text(75,28.5)[]{$2$} \Text(110,28.5)[]{$1$} \Text(45,10.5)[]{$4$} \Text(45,46.5)[]{$3$}
    \end{picture}}
  \caption{{\upshape (a)} The assignment between the geometrically constructed cycles between
    D-branes and the cycles that are given in the table above. This assignment is the same for
    each of the four $SO(8)$ blocks. The cross marks the position of the O~plane, grey dots
    denote the D7~branes. Due to the fibre involution in the O-plane monodromy, cycles 3 and 4 do
    not intersect. {\upshape (b)} The corresponding gauge enhancement: The cycles become the nodes
    of the Dykin diagram, lines are drawn for intersections. 
    % Note that the distribution of the cycles 1,3 and 4 is ambiguous.
    % XXX distribution ambiguous = order arbitrary ? = Triality?XXX 
    \label{1block}}
\end{center}
\end{figure}

These cycles can be constructed geometrically. Their projections to the base $\mathbb{C}\Pp^1$
connect the D-branes and are displayed for one block in Figure~\ref{1block}.

We can now move away from the $SO(8)^4$ configuration by rotating $\omegat$. A convenient parameterisation is
given by 
\begin{align}
\omegat=\frac{1}{2}\left(\alpha+Ue_2+S\beta-\left(US-z^2\right)
e_1+2\widehat{E}_{I}z_I\right) \ ,\label{Omegagen}
\end{align}
with shifted $E_8\times E_8$ block vectors  $\widehat{E}_I= E_I +W_I^1 e_1 +W_I^2 e_2$. 
Explicitly, they are
\begin{align}\label{eq:Ehat}
  \begin{aligned}
    \widehat{E}_1&=E_1+e_2 ,\hspace{1cm} \widehat{E}_{I}=E_I,&& I=2..4,
    10..12 \,,\\
    \widehat{E}_9&=E_9+e_2 ,\hspace{1cm} \widehat{E}_{J}=E_J+e_1/2,&&
    J=5..8, 13..16 \, . 
  \end{aligned}
\end{align}
The $\widehat{E}_I$ are orthogonal to $\alpha$ and $\beta$ and still satisfy $\widehat{E}_I\cdot
\widehat{E}_J=-\delta_{IJ}$. Using Table~(\ref{AtoD}) and Figure~\ref{1block}, one can show \cite{bht08} that
the $z_I$ are the positions of the branes relative to their respective O-planes in the double cover
of  $\mathbb{C}\Pp^1\simeq T^2/Z_2$. 

Now we can deduce the brane positions and the gauge enhancement from
a given expansion of the holomorphic two-form $\omegat$ (which is equivalent to knowing
the complex structure of $\widetilde{K3}$).
We can either match any expansion of $\omegat$ in the basis given in Appendix~\ref{sec:K3} to
(\ref{Omegagen}), or we can compute the intersection numbers between $\omegat$ and the cycles given
in Table~$\ref{AtoD}$ to find the periods of the cycles of $\widetilde{K3}$. In this way we obtain the
value of the dilaton and the D-brane and O-plane positions. Note that contrary to the
basis given by $\alpha, \beta, e_1, e_2$ and the cycles in Table~(\ref{AtoD}), the basis we used in
the expansion~(\ref{Omegagen}) is not an integral basis (as the $\widehat{E}_I$ are half-integral).

% In the next subsection, we will describe a systematic procedure for determining a flux that fixes a given D7-brane configuration.
% In the following subsections, we will apply this procedure: We will show how to stabilise the complex structure in a $SO(8)^4$ configuration,
% i.e.~all $z_I=0$. We will present three further examples: two in which we break one of the four
% $SO(8)$ factors to $SO(6)$ and one in which we break it to $SO(4)\times SU(2)$, corresponding to
% moving one or two branes off the O-plane, respectively.

\vskip 8mm

\subsection[Fixing D7-brane Configurations by Fluxes]{Fixing D7-brane Configurations by Fluxes}

We are now ready to outline a systematic procedure for choosing a flux which fixes a given D7-brane gauge group.
In particular, we will be interested in non-Abelian gauge enhancement. The Cartan matrix of the underlying Lie-Algebra
is given by the intersection matrix of the lattice of shrinking two-cycles. Thus, we need to understand which fluxes
make a particular subspace of two-cycles shrink.
We will take these cycles as part of the basis orthogonal to $\left<\widetilde{B},\widetilde{F}\right>$ discussed at the end of Section \ref{sec:modstab}. Then we consider the orthogonal lattice, i.e. the lattice made up of (integral) cycles orthogonal to the shrinking ones (and to $\left<\widetilde{B},\widetilde{F}\right>$). Choosing an integral basis for this lattice completes the basis of cycles of $H_2(\Kt)$ orthogonal to $\left<\widetilde{B},\widetilde{F}\right>$. Note that in this basis the metric on $H_2(\Kt)$ is block-diagonal, with a negative definite block for the subspace of shrinking cycles. We also choose a basis of integral cycles of $H_2(K3)$ such that the metric has two blocks with the same dimensions as on the $\Kt$ side.

In this basis it is easy to write down a flux that fixes $\omegat$ orthogonal to the shrinking cycles: It can be taken to have the block-diagonal form\footnote{Actually, it is enough that $G^aG$ is of this form.}
\begin{equation}\label{Gprocedure}
\left(\begin{array}{cc}
 G_\perp &  \\ & G_{\rm shk}  \\ 
\end{array}\right)\:.
\end{equation}
Thus, when diagonalising $G^aG$, the positive norm eigenvectors are in the first block and hence orthogonal to the shrinking cycles.

One has finally to check whether there are more shrinking cycles than those we imposed.

\subsection[Fixing an $SO(8)^4$ Point]{Fixing an \boldmath$SO(8)^4$ Point\label{sec:so8stab}} 
In this section, we will follow the procedure described in the previous section to construct
a flux that fixes the F-theory moduli corresponding to four
D7 branes on top of each O7 plane. This $SO(8)^4$ configuration is realised when there are
sixteen shrinking cycles whose intersection matrix is $D_4^4$. These shrinking cycles are given by the four blocks
$A,B,C,D$ as defined in~(\ref{AtoD}). The basis of the orthogonal lattice is given by
$\alpha,e_1,\beta,e_2$ (see Eq.~(\ref{eq:alphabeta})). Since the only
nonvanishing intersections in this set are $\alpha\cdot e_1=\beta\cdot e_2=2$, the
intersection matrix is
\begin{align}
  \Mt = \begin{pmatrix}
    \begin{matrix}
      0&2\\2&0\\
    \end{matrix} &&\\
    &\begin{matrix}
      0&2\\2&0\\
    \end{matrix}&\\
    && D_4^4
  \end{pmatrix}\,.
\end{align}
For $K3$ we choose the same basis. Note that we are ignoring the $U$
block spanned by base and fibre.

Then we take the $20\times 20$ flux matrix %(again ignoring the F-theory zero $2\times 2$ block)
with respect to these bases to be
\begin{align}\label{eq:so8flux}
G^{I\Lambda}=\left(\begin{array}{ccc}
G_1 &&\\ &G_2 &\\ && \mathbf{0}_{16}\\
\end{array}\right)\,,
\end{align}
where $G_1$ and $G_2$ are $2\times 2$ blocks (which form the $G_\perp$ of \eqref{Gprocedure})
and the zero block is $16\times 16$ ($G_{\rm shk}$ of \eqref{Gprocedure}). If $G_1$ and $G_2$
satisfy the condition to have minima, then one $\omegat_j$ is fixed along the space
$\left<\alpha,e_1\right>$,  while the other is fixed in the space $\left<\beta,e_2\right>$. This
immediately gives a complex structure $\omegat$ that is orthogonal to the $D_4^4$ blocks
$A,B,C,D$ and hence realises an $SO(8)^4$ point.

An explicit example of an integral flux that satisfies the tadpole cancellation condition
(tr$\,G^aG=48$) and fixes an $SO(8)^4$ point is given by: 
\begin{align}\label{G1G2so8}
  G_1 = \left(\begin{array}{cc}
      1&1\\1&1\\
    \end{array}\right)\,, \qquad\qquad\qquad
  G_2 = \left(\begin{array}{cc}
      1&1\\1&3\\
    \end{array}\right)\,.
\end{align}
The corresponding blocks for $G^aG$ are
\begin{align}
  (G^aG)_1 = \left(\begin{array}{cc}
      8&8\\8&8\\
    \end{array}\right)\,, \qquad\qquad\qquad
  (G^aG)_2 = \left(\begin{array}{cc}
      16&24\\8&16\\
    \end{array}\right)\,,
\end{align}
and the corresponding eigenvalues are
\begin{align}\label{eigenvalSO(8)4}
  \lambda_{\omegat_1} = 16\,, \qquad \lambda_{\tilde{u}_1} = 0\,,  \qquad\
  \lambda_{\omegat_2} = 8(2+\sqrt{3})\,,  \qquad \lambda_{\tilde{u}_4} =
  8(2-\sqrt{3})\,.
\end{align}
We see that their sum is precisely $48$, as required by tadpole cancellation, and that they are all
non-negative, as required by the minimum condition. Moreover, the ones corresponding to positive norm
eigenvectors are different from those relative to negative norm eigenvectors, as required by the
stabilisation condition. 

The positive norm eigenvectors of the two matrices give $\omegat_1$,
$\omegat_2$:
\begin{align}
 \omegat_1 &= \frac{\alpha}{2} +\frac{e_1}{2}\,,
 &\omegat_2 &=  3^{1/4}\,\frac{\beta}{2} +\frac{1}{3^{1/4}}\,\frac{e_2}{2}\,.
\end{align}
From the comparison of $\omegat=\omegat_1 + \I \omegat_2$ with the general form~(\ref{Omegagen}), we
see that indeed the complex structure is fixed at a (non-integral) point where $z_I=0$, and that the complex
structures of base and fibre are given by
\begin{align}
  U&= \frac{1}{\sqrt[\leftroot{2}4]{3}}\, \I \,, & S&=\sqrt[\leftroot{2}4]{3} \,\I \,.
\end{align}
Since $S$ is the type IIB axiodilaton, we have stabilised the string coupling at a moderately small
value of $3^{-1/4}\cong 0.76.$ However, we can probably realise smaller coupling by considering
generic $4\times 4$ matrices rather than the $2\times 2$ block structure of Eq.~(\ref{eq:so8flux}).  

This flux fixes also the deformations of $\omega_1$ and $\omega_2$. On the other hand,
$\omega_3$ and $\omegat_3$ are eigenvectors of $G\,G^a$ and $G^aG$
relative to zero eigenvalues. Then their deformation along all negative
eigenvectors relative to zero eigenvalues are left unfixed.
In type IIB, this corresponds to leaving unfixed K\"ahler moduli of
$K3\times
T^2/\mathbb{Z}_2$, while fixing the complex structure and the D7-brane
positions. The unfixed deformations of $\omegat_3$ correspond to gauge
fields in type IIB that remain massless \cite{v08}.
In the studied case, two of the $19\times 2$ deformations of $\omega_3$ and $\omegat_3$, the ones
along $\tilde{u}_4$, are fixed (as $\lambda_{\tilde{u}_4} = 8(2-\sqrt{3})$ is different from zero). Fixing a
deformation of $\omegat_3$ corresponds to giving a mass to the corresponding
gauge field in type IIB dual. In fact, this flux corresponds to the type IIB flux
that makes one four-dimensional vector massive \cite{aaf03,aaft03,hkl06,jl04}. One can see this also from the M-theory point of view: One three-dimensional vector gets a mass from fluxes. This vector combines with the deformation of $\omegat_3$ to give a four-dimensional massive vector.

Finally we note that the lower $K3$ is generically non-singular, as $\omega_3$
will generically not be orthogonal to the $E_8$ block cycles.

\

As a second example we will reproduce one of the solutions given in \cite{ak05}
by using our methods. As it is discussed there, attractive $K3$ surfaces are classified in terms of
a matrix
\begin{equation}
 Q = \left(\begin{array}{cc}
      p\cdot p&p \cdot q\\p \cdot q&q \cdot q\\
    \end{array}\right)\, ,
\end{equation}
in which $p$ and $q$ are integral two-forms. 
The holomorphic two-form of $\Kt$ is then given by
\begin{equation}
\omegat=\pt+\tau \qt \,.\label{Ompq}
\end{equation}
Of the 13 pairs of attractive $K3$'s given in \cite{ak05}, we will
discuss the one defined by
\begin{equation}
 Q = \left(\begin{array}{cc}
      8 & 4 \\ 4 & 8 \\
     \end{array}\right),\qquad \qquad
\widetilde{Q}=\left(\begin{array}{cc}
      4 & 2 \\ 2 & 4 \\
    \end{array}\right)\, .\label{QtildeQ}
\end{equation}
This pair has the advantage that both $K3$'s have an orientifold interpretation
which means that we can expand $\pt$ and $\qt$ in terms of $e_1$ , $e_2$, $\alpha$ and $\beta$
(and similarly, for the lower $K3$, $p$ and $q$ in terms of $e_1'$ , $e_2'$, $\alpha'$ 
and $\beta'$). Clearly, there are many ways to do this which correspond to different embeddings
of the lattice spanned by $p$ and $q$ into the lattice spanned by $e_1$ , $e_2$, $\alpha$ 
and $\beta$. We make the following choice:
\begin{align}
  \begin{aligned}
    p=&e'_1+2\alpha'+2\beta'\,,   &\tilde{p}=&e_1+\alpha+\beta\,,\\
    q=&e'_2+2\beta'	 \,,  &\tilde{q}=&e_2+\beta\, .\label{pqexp}
  \end{aligned}
\end{align}
According to \cite{ak05}, stabilization at this point occurs through the flux
\begin{equation}
G=\frac{1}{2}\left(\gamma\omega\wedge\overline{\omegat}+\overline{\gamma}\,\overline{\omega}
\wedge\omegat\right)
\end{equation}
with $\gamma=1+\frac{i}{\sqrt{3}}$. In the basis given by $\alpha,e_1,\beta,e_2$ and $\alpha',e'_1,\beta',e'_2$,
%$\tilde{\alpha},\tilde{e}_1,\tilde{\beta},\tilde{e}_2$, 
the flux matrix reads
\begin{align}
 G^{I\Lambda}= \begin{pmatrix}
    2 &  2 & 2 & 0   \\
    1 & 1 & 1 & 0  \\
    0 &  0 & 2 & 2 \\
    -1 & -1  & 0 & 1
  \end{pmatrix}\, .
\end{align}

The positive norm eigenvectors of $G^aG$ are given by $\omegat_1=(1,1,\frac{1}{2},-\frac{1}{2})$ and 
$\omegat_2=(0,0,1,1)$. Rescaling the second one so that they both have the same
norm, we arrive at $\tilde{\omega}=\omegat_1+i\frac{\sqrt{3}}{2} \omegat_2$. This is precisely the same result as 
what one obtains from inserting (\ref{pqexp}) into (\ref{Ompq}). 

The eigenvalues of $G^a G$ are $\lambda_{\omegat_1} = \lambda_{\omegat_2}=24$, $\lambda_{\tilde{u}_1}=\lambda_{\tilde{u}_2}=0$. In the last section we will see that this corresponds to an $\Nn=1$ (4d) vacuum. Moreover, in this case all the K\"ahler moduli of both $K3$'s are left unfixed by fluxes, as all the eigenvalues $b_a$ are equal to zero.

\subsection{Moving  Branes by Fluxes\label{sec:so6stab}}

Now we want to see how to change the flux~\eqref{eq:so8flux}, with $G_1$ and $G_2$ given by
\eqref{G1G2so8}, to fix a different D7-brane configuration in which some D7 branes have been moved
away from the orientifold planes. In particular, we will find fluxes that fix configurations
where we move one or two branes off one of the stacks, breaking one $SO(8)$ to
$SO(6)$ or $SO(4)\times SU(2)$. In the following we will consider only the $C$-block.
The cycles belonging to blocks $A,B,D$ will remain shrunk.

\subsubsection*{\boldmath$SO(8)^3\times SO(6)$}
\label{1stExampleBrMv}
Moving one D7~brane from one stack in type IIB corresponds to blowing up one of the 4 cycles of this block. % this is achieved by
%blowing up one of the 16 cycles %$\widetilde{A}_h,\widetilde{B}_h,\widetilde{C}_h,\widetilde{D}_h$ 
%$A_h,B_h,C_h,D_h$.%, as explained at page \pageref{1stExampleBrMv}.
For the first example, % \label{1stExampleBrMv} 
consider the complex structure determined by~(\ref{Omegagen}) with $z_1=d$ and al other $z_I=0$.
One can check that all cycles given in Table~\ref{AtoD} except $C_1$ remain orthogonal to $\omegat$.
Looking at Figure~\ref{1block}, it is clear that this means we have moved one D-brane away from the
O-plane, as claimed. Thus $SO(8)$ is broken to $SO(6)$.
At the same time the cycles that remain shrunk in block $C$ have an intersection matrix that
is equivalent to minus the Cartan matrix of $SO(6)$. This means that we have effectively
crossed out the first line and the first column of the Cartan matrix of $SO(8)$ by removing
$C_1$ from the set of shrunk cycles:
\begin{align}
  \begin{pmatrix}
    -2 & 1 & 0 & 0   \\
    1 & -2 & 1 & 1  \\
    0 &  1 & -2 & 0 \\
    0 & 1  & 0 & -2
  \end{pmatrix}
  \longrightarrow 
  \begin{pmatrix}
    -2 & 1 &1  \\
    1 & -2 & 0 \\
    1  & 0 & -2 
  \end{pmatrix}\,.
\end{align}
We want an integral basis in which shrunk and blown-up cycles do not intersect each 
other. %However we still want the basis cycles to be orthogonal to $\omegat$ at an $SO(8)$ point. 
To achieve this we keep the shrunk cycles $C_2$, $C_3$, $C_4$ and instead of $C_1$
we take the integral cycle $2\widehat{E}_1=2\left(e_2+E_1\right)$ (see (\ref{eq:Ehat})) to describe the
brane motion in block $C$.
We find the intersection matrix    
\begin{align}\label{1stExampleIntMatrix}
  \begin{pmatrix} 
    -4 & 0 & 0 & 0  \\
    0 & -2 & 1 & 1  \\
    0 &  1 & -2 & 0 \\
    0 & 1  & 0 & -2
  \end{pmatrix}\,. 
\end{align} 
We choose an analogous basis for the lower $K3$.

\

The basis $\alpha,e_1,\beta,e_2,2\widehat{E}_1,C_2,C_3,C_4,A,B,D$, is the one that gives the flux matrix the block-diagonal form \eqref{Gprocedure},
with the shrinking cycles given by $C_2,C_3,C_4,A,B,D$ and the orthogonal ones by $\alpha,e_1,\beta,e_2,2\widehat{E}_1$.
%integral, and so any
%integer $G^{I\Lambda}$ gives an integral flux. Clearly, the flux that stabilises this situation
%cannot be block-diagonal as in the $SO(8)^4$ case, (\ref{eq:so8flux}), but has to 
Such a block-diagonal flux matrix generally gives $\omegat$ a component along
$\widehat{E}_1$. An example is given by: % suitable flux that fixes such an $\omegat$ is
\begin{align}
  G^{I\Lambda} = \begin{pmatrix}
    1 & 1 & & & & \\
    1 & 1 & & & & \\
    & & 1 & 1 & 0 & \\
    & & 1 & 3 & 1 & \\
    & & 0 & 1 & 0 &\\
    & & & & & \mathbf{0}_{15}\\
\end{pmatrix}\,.
\end{align}
where the $3\times 3$ block is with respect to the cycles $\beta,e_2,2\widehat{E}_1$ for
both $K3$'s. From the type IIB perspective, we are also turning on fluxes on the D7~branes.

$G^aG$ satisfies the tadpole cancellation condition. The eigenvalues corresponding to the first
block are the same as in Eq.~(\ref{eigenvalSO(8)4}), the ones in the second block are   
\begin{align}
  \lambda_{\omegat_2} = 24.6\,, \qquad\ \lambda_{\tilde{u}_2} = 5.5\,,  \qquad \lambda_{\tilde{u}_4}
  = 1.9 \,. 
\end{align}
They are all positive and different from each other. The positive norm eigenvectors give
$\omegat_1$ and $\omegat_2$: 
\begin{align}
 \omegat_1 = \frac{\alpha}{2} + \frac{e_1}{2} \qquad,\qquad \omegat_2 = 0.9 \,\frac{\beta}{2} + 1.3
 \frac{e_2}{2} + 0.3\, \widehat{E}_1\:.
\end{align}
The corresponding $\omegat$ is orthogonal the $S^2$ cycles with intersection matrix
$SO(6)\times SO(8)^3$, but it is not orthogonal to the cycle $2\widehat{E}_1$ which is now blown up, at a
volume $\rho\!\left(2\widehat{E}_1\right)=0.6\,\sqrt{\nut}$. This
corresponds to the motion of one D7~brane away from the orientifold plane of block $C$. 
Note that again the coupling is moderately weak, $g=1/1.61 = 0.6$. 

We also note that, with respect to our $SO(8)^4$-example, we have fixed one more deformation
of $\omega_3$ and one of $\omegat_3$.
%The first one corresponds to the fact that D7-brane fluxes can fix K\"ahler moduli.
The stabilisation of the extra
$\omegat_3$ deformation is the signal of a mass for the gauge field on the D7~brane that has been
moved. This mass is explained in type IIB by the fact that D7 fluxes gauge some shift symmetries by
vectors on the branes. Since the $U(1)$ on the brane is broken, the resulting gauge group is
$SO(8)^3\times SO(6)$ \cite{aaf03,aaft03,hkl06,jl04}.

\subsubsection*{\boldmath$SO(8)^3\times SO(6)\times U(1)$}
\label{1stExampleBrMvbis}
In the example studied above, we have given a flux that fixes the desired brane configuration.
Moreover it fixes one further deformation of $\omega_3$ and one of $\omegat_3$, with respect to 
the $SO(8)^4$ example presented before. This is related to the fact
that the rank of the $3\times 3$ block has been increased to $3$; so we get two negative norm eigenvectors with
non-zero eigenvalues. But we can choose a different flux, such that the number of negative
norm eigenvectors relative to non-zero eigenvalues does not change with respect to the $SO(8)^4$
case:
\begin{align}
  G^{I\Lambda} = \begin{pmatrix}
    1 & 1 & & & & \\
    1 & 1 & & & & \\
    & & 1 & 1 & 0 & \\
    & & 1 & 3 & 1 & \\
    & & 0 & 0 & 0 &\\
    & & & & & \mathbf{0}_{15}\\
\end{pmatrix}\:,
\end{align}
where the $3\times 3$ block is still with respect to the cycles $\beta,e_2,2\widehat{E}_1$.
% for $\widetilde{K3}$ and for three integral cycles of $K3$ that have the same intersection matrix.

Again, $G^aG$ satisfies the tadpole cancellation condition. The eigenvalues relative to the first block are the same as in
Eq.~(\ref{eigenvalSO(8)4}). The eigenvalues of the second block are
\begin{align}
  \lambda_{\omegat_2} = 27.3\:, \qquad\ \lambda_{\tilde{u}_2} = 4.7 \:, \qquad \lambda_{\tilde{u}_4} = 0\:.
\end{align}
They are all non-negative and different from each other. The positive norm eigenvectors give
$\omegat_1$ and $\omegat_2$: 
\begin{align}
 \omegat_1 = \frac{\alpha}{2} + \frac{e_1}{2} \qquad,\qquad \omegat_2 = 0.8 \,\frac{\beta}{2} + 1.4
 \frac{e_2}{2} + 0.3\, \widehat{E}_1 \:.
\end{align}
As before, the corresponding $\omegat$ is orthogonal the $S^2$ cycles with intersection matrix
$SO(6)\times SO(8)^3$, but it is not orthogonal to the cycle $2\widehat{E}_1$ which is now blown up, at a
volume $\rho\!\left(2\widehat{E}_1\right)=0.6\,\sqrt{\nut}$. Again, one D7~brane is moved from the orientifold plane of block $C$. 
%Note that again the coupling is moderately weak, $g=1/1.61 = 0.6$. 

In this case, we do not break any further $U(1)$. In fact, the flux we
turned on contributes to the gauging of an isometry that has been
gauged also in the $SO(8)^4$ case. This can be easily understood in the M-theory context, where the relevant gauge field is one of the $\Ct^\Lambda_{1\mu}$.

% We finally note that in this case the negative norm eigenvector relative to the zero eigenvalue is
% parallel to an integral vector.

\subsubsection*{\boldmath$SO(8)^3\times SO(4)\times SU(2)$}
As a further example\label{2ndExampleBrMv}, let us choose $z_1=z_2=d$ and all other $z_I=0$. We
now find that $\omegat\cdot C_2=d$. For all other cycles in
Table~(\ref{AtoD}) the 
intersection with $\omegat$ still vanishes, so we have blown up a different cycle than in the
previous examples. From the assignment between cycles and forms it is clear that we have moved two
branes away from the O-plane. As $C_1$ remains shrunk,
these branes are on top of each other. From the type IIB perspective, we thus expect the
gauge symmetry $SO(4)\times SU(2)$. Examining the intersection matrix of the shrunk cycles $C_1$,
$C_3$ and $C_4$ we indeed find a diagonal matrix with entries $-2$. This happens because we have
blown up the cycle $C_2$ and thus deleted the second row and second column from the Cartan matrix
of
$SO(8)$: 
\begin{align}
  \begin{pmatrix}
    -2 & 1 & 0 & 0   \\
    1 & -2 & 1 & 1  \\
    0 &  1 & -2 & 0 \\
    0 & 1  & 0 & -2 
  \end{pmatrix}
  \longrightarrow 
  \begin{pmatrix}
    -2 & 0 & 0\\
    0 & -2 & 0 \\
    0 & 0  & -2 
  \end{pmatrix}\,.
\end{align}
The result is minus the Cartan matrix of $SO(4)\times SU(2)$, as expected.
As before, we need a basis of integral cycles in which shrunk and blown-up
cycles do not intersect. % and which is still orthogonal to $\omegat_{SO(8)^4}$. 
To construct it, we replace the cycle $C_2$ with the cycle $\widehat{E}_1+\widehat{E}_2= e_2+E_1+E_2$. It has
self-intersection $-2$, so that the intersection matrix in the new basis of cycles which we use for D-brane motion 
in the $C$ block is 
\begin{align}\label{2ndExampleIntMatrix}
  \begin{pmatrix}
    -2 & 0 & 0 & 0   \\
    0 & -2 & 0 & 0  \\
    0 &  0 & -2 & 0 \\
    0 & 0  & 0 & -2 
  \end{pmatrix}\,.
\end{align}
In this basis, a flux that stabilises the desired gauge group is given by:
\begin{align} 
  G^{I\Lambda} &= 
  \begin{pmatrix}
    1&1 &&&& \\
    1&1 &&&&\\ 
    && 1& 1& 1 &\\
    &&1&3&1\\
    &&1&1&2&\\
    &&&&& \mathbf{0}\\
  \end{pmatrix}\,,
\end{align}
where now the $3\times 3$ block is with respect to the cycles $\beta,e_2,\widehat{E}_1+\widehat{E}_2$. The
eigenvalues corresponding to this block are: 
\begin{align}
  \lambda_{\omegat_2} = 19.6\,, \qquad\ \lambda_{\tilde{u}_2} = 11.2\,,  \qquad \lambda_{\tilde{u}_4}
  = 1.2\,.
\end{align}
They are all positive and different from each other. $\omegat_1$ and $\omegat_2$ are given by:
\begin{align}
  \omegat_1 &= \frac{\alpha}{2} + \frac{e_1}{2} \,,& \omegat_2 & =  1.5\,\frac{\beta}{2} +
  0.8\,\frac{e_2}{2} - 0.3\, \left(\widehat{E}_1+\widehat{E}_2\right)\:.
\end{align}
The corresponding $\omegat$ is orthogonal the $S^2$ cycles with intersection matrix
$SO(4)\times SU(2) \times SO(8)^3$, but it is not orthogonal to the cycle
$\widehat{E}_1+\widehat{E}_2$ which is now blown up. 

Also in this example, we have fixed one further deformation of $\omega_3$ and one of $\omegat_3$.
This in particular breaks the gauge group on the two D7~branes from $U(2)$ to $SU(2)$.

\subsection{Fixing almost all Moduli}
In the previous examples we have considered fluxes that stabilise the D7-brane positions and part of the
metric moduli of $K3$, while leaving some geometric moduli unfixed. This was due to the large amount
of zero eigenvalues of $G^aG$. In what follows, we will present an example of an integral flux that
satisfies the tadpole cancellation condition and fixes almost all geometric moduli. The
remaining unstabilised moduli are the size of the fiber in $\Kt$, as prescribed by the F-theory
limit, three deformations of $\Sigma$, and the two volumes of $K3$ and $\Kt$.  

To write down the flux we will choose two different bases of integral cycles in $H^2(K3)$ and in $H^2(\Kt)$. The
second one is the same as in the example $SO(8)^4$, while for $H^2(K3)$ we choose an integral basis
with intersection matrix %$\clubsuit$write explicitly the BASIS?$\clubsuit$
\begin{equation}
\left(\begin{array}{cccc}
\begin{array}{cc} 0&1\\1&0\\ \end{array} &&& \\
&\begin{array}{cc} 0&1\\1&0\\ \end{array} && \\
&&\begin{array}{cc} 0&1\\1&0\\ \end{array} & \\
&&& D_4^4\\
\end{array}\right)\,.
\end{equation}
In these bases, we choose the flux matrix to be
\begin{equation}
G^{I\Lambda}=\left(\begin{array}{ccc}
%\begin{array}{cc} 0&0\\0&0\\ \end{array} &&& \\
\begin{array}{rr} 1&-1\\-1&1\\ \end{array} && \\
&\begin{array}{rr} 1&-1\\-1&1\\ \end{array} & \\
&& G_{(4)}^4\\
\end{array}\right)\,.
\end{equation}
where 
\begin{equation}
G_{(4)}=\left(\begin{array}{rrrr}
-1 & -1 &0 &0 \\ 0 & 0 & 1 & 1 \\ 0 & 0 & 1 & 0 \\ 0 & -1 & 0 & 0 \\ 
\end{array}\right) \:.
\end{equation}

This flux satisfies tr$\,G^aG=8+8+4\times 8 =48$. Moreover, the $G_{(4)}$ blocks have eigenvalues equal
to 2, while the $2\times 2$ blocks have eigenvalues equal to $0$ for the positive norm eigenvectors
and $8$ for the negative norm eigenvectors. In the next section, we will see that the resulting minimum is supersymmetric ($\Nn=2$ in 4d). The eigenvalues relative to positive norm eigenvectors are such that all moduli are fixed apart from the
deformations of the $\omega_i$'s and the $\omegat_j$'s in the first U-block\footnote{This is a singular example, as now the lower $K3$ is singular.}.

\section{SUSY Vacua\label{sec:susyvac}}

Finally, we want to study the question of supersymmetric vacua. This question has been analysed
for M-theory on a generic eight-dimensional manifold in~\cite{bb96,drs99}.
In the presence of fluxes a supersymmetric solution
is a warped product of $\mathbb{R}^{1,2}$ and some internal manifold which is conformally
Calabi--Yau \cite{bb96}. %In the following, we will consider the K\"ahler form and 
%complex structure of the underlying Calabi--Yau manifold.
The flux $G_4$ must be primitive ($J\wedge G_4=0$) and of Hodge type $(2,2)$ with respect to the K\"ahler form and the
complex structure of the underlying Calabi--Yau\footnote{
In the following, all the quantities of the internal manifold are relative to the unwarped Calabi--Yau metric.
}. Given a metric with $SU(4)$ holonomy, there is only one associated K\"ahler form $J$
and one  holomorphic four-form $\Omega$. Moreover there are only two invariant Majorana--Weyl spinors, which implies $\Nn=2$ supersymmetry in the three-dimensional theory.

In our case, $K3\times K3$ has holonomy $SU(2)\times SU(2)$. As we have seen previously, for each $K3$ factor, 
the metric is invariant under the $SO(3)$ that rotates the $\omega_i$'s. This means that, given the metric
of $K3\times K3$, there is an $S^2\times S^2$ of
possible complex structures and associated K\"ahler forms.
Moreover, the holonomy $SU(2)\times SU(2)$ implies that the number of globally defined Majorana--Weyl spinors is four,
corresponding to $\Nn=4$ supersymmetry in three dimensions. The $R$-symmetry is the $SO(4)\simeq SO(3)\times SO(3)$
that rotates the four real spinors and the corresponding $S^2\times S^2$ of complex structures.
When this symmetry is broken to the $SO(2)$ which rotates the real and imaginary part of $\Omega$,
then we have $\Nn=2$ supersymmetry. On the other hand, if it is completely broken we have $\Nn=0$.

A minimum is supersymmetric if we can associate with the metric a K\"ahler form $J$ and a complex structure $\Omega$, such that $G_4$ is primitive and of Hodge-type (2,2). This means that there must be a choice of $\omega_i$ and $\omegat_j$, let us say $J=\sqrt{2\nu}\,\omega_3+\sqrt{2\nut}\,\omegat_3$ and  $\Omega=\omega\wedge\omegat$ (with $\omega = \omega_1+\I \omega_2$ and $\omegat = \omegat_1+ \I \omegat_2$), such that $G_4 \wedge J =0$ and $G_4\wedge \Omega =G_4\wedge \bar{\Omega} =0$. In our formalism, this is equivalent to:
\begin{itemize}
  \item Primitivity, $G_4 \wedge J = 0$ :
    \begin{align}\label{primitCond}
      G\,\omegat_3&=0\,, & G^a \omega_3&=0\,.
    \end{align}
    In terms of the eigenvalues of $G^aG$ this means $a_3=0$. We see that the primitivity condition translates to the existence of a non-trivial kernel of $G^aG|_{\Sigmat}$ and $G\,G^a|_{\Sigma}$. The vectors in the kernels make the K\"ahler form.
  \item $G_4=G_4^{(2,2)}$:
    \begin{equation}\label{22Cond}
	0 = (\omega \cdot G\omegat) = a_1 - a_2 \,\,\,.
    \end{equation}
    This means $a_1=a_2\equiv a$.
\end{itemize}

To summarise, the necessary and sufficient condition for the flux to preserve susy in the minimum is that 
$G$ (when restricted to the block $\Sigmat,\Sigma$) takes the form
\begin{align}
  G\big|_{\Sigmat}&=\begin{pmatrix} a&&\\&a&\\&&0\end{pmatrix} \,.
\end{align}
For $a=0$, the $SO(4)$ $R$-symmetry is unbroken and the minimum preserves all the $\mathcal{N}=4$ supersymmetries.
For $a\neq0$, only an $SO(2)$ subgroup of the $R$-symmetry is preserved
and we have $\mathcal{N}=2$ supersymmetries in three dimensions.

We note that in the case of fluxes which are compatible with the F-theory limit, the condition
$a_3=0$ is always satisfied and so one has simply to check that the other two eigenvalues are equal
to each other or possibly zero.

\section{Conclusions\label{sec:conclusions}}

In this paper, we have analysed in detail the stabilisation of D7-brane configurations by fluxes.
To do that we have used the F-theory language, i.e.\ we have studied the stabilisation problem
in M-theory and then mapped the results to type IIB.

We studied the stabilization of D7/O7 configurations on $K3\times
T^2/\mathbb{Z}_2$. The  O7~planes and the D7~branes are wrapped on $K3$ and localised on
$T^2/\mathbb{Z}_2$; in particular, the O-planes sit at the four singularities of $T^2/\mathbb{Z}_2$.
The D7 moduli are the positions of the D7~branes on $T^2/\mathbb{Z}_2$. The M-theory dual of this
background is given by the compactification on $K3\times K3$ (in the F-theory limit), where the
second $K3$ is elliptically fibred.

Our aim was to analyse the moduli stabilisation, in this background, by integral three-form closed
string fluxes and by D7 worldvolume two-form fluxes, using F-theory language. The type IIB geometric
and D7 moduli are all mapped to M-theory geometric moduli. Three-form and two-form fluxes are
both mapped to four-form fluxes.

We have considered M-theory on $K3\times K3$ and derived the four-form flux generated potential
for the geometric moduli in the Section~\ref{sec:fluxpotential}. We have expressed it in terms of the
three orthogonal vectors of $H^2(K3)$ that determine the metric of $K3$. Furthermore, we explictly found the flux-induced mass terms for the
vector fields coming from the three-form field. In the Section~\ref{sec:modstab} we have worked out in detail the
moduli stabilisation, finding the geometric conditions for a flux to
minimise the potential: It must map the three-plane of one $K3$ to the three-plane of the other $K3$
and back. Using the duality, we can map the stabilised point found in M-theory moduli space to a point in type IIB
moduli space. In this way we can see which D7 configuration is stabilised by a particular flux.  

The M-theory fluxes dual to Poincar\'e-symmetry-preserving type IIB fluxes do not stabilise the size of the
fibre. So we always have a flat direction in the M-theory moduli space. Of this line, only one point
corresponds to a four dimensional vacuum, the one associated with zero fibre size. We have verified
that it is at infinite distance from any other point in the moduli space. The F-theory
limit consists in  going to this specific point along the flat direction. %We can do that because the direction in which we take
%the limit is never stabilised.
We have described this limit in detail in Section~\ref{sec:ftheory}.
In particular, we have seen which moduli disappear from the M-theory moduli space when
we take the F-theory limit. %We have identified these moduli with the deformations of $\omegat_3$.

In Section~\ref{sec:movebrain} we have studied some examples. First, we have reviewed the map
between the D7 moduli and the dual M-theory geometric dual moduli worked out in
\cite{bht08}. This map enabled us to outline an explicit procedure to find a flux that stabilises
a desired gauge group via its pattern of shrinking cycles. Using this procedure, we have shown a
flux that stabilises 4~D7~branes on top of
each O-plane. Then we have found which fluxes we have to turn on to modify this configuration
and move one or two branes away from one O-plane. This changes the gauge group in type IIB.
Correspondingly, the flux fixes a different singularity in the upper $K3$, i.e.\ some cycles are blown up.
% In particular, we have found the condition to blow up some cycles of the upper
% $K3$: They must be mapped by the flux matrix to cycles with positive self-intersection in the lower
% $K3$. In our example we have switched on the components of the flux matrix that precisely do that. 

In the examples we have also checked whether there are some stabilised K\"ahler moduli of the lower $K3$.
When this is the case, some K\"ahler moduli of the upper $K3$ are stabilised too. These are mapped to
the fourth components of four-dimensional vector fields \cite{v08}. The corresponding three-dimensional scalars
acquire a mass since they are stabilised. The corresponding three-dimensional vectors also
become massive (see Section \ref{sec:vectormass}). So we concluded that the resulting
four-dimensional vectors acquire a mass. This result matches with what was found in \cite{aaf03,aaft03}, studying
directly type IIB on $K3\times T^2/\mathbb{Z}_2$ (see also \cite{jl04,hkl06}).

At the end of Section \ref{sec:movebrain}, we have reported one further example. We have presented a
flux that stabilises almost all the moduli, showing that a general F-theory flux would fix almost all the
moduli (except one K\"ahler modulus in the lower $K3$, that corresponds to the fibre size in the upper $K3$).

In the last section we have considered the sypersymmetry conditions on the set of the four-dimensional
Minkowski vacua we have studied. In general supersymmetry is completely broken, but under
some conditions, the $\Nn=1$ or even $\Nn=2$ supersymmetry in four dimensions can be preserved.
We have found these conditions using an eleven-dimensional approach.

\

In this work we have studied a particular example, $K3\times K3$, in which we have complete control over D7-brane stabilisation by fluxes.
This is due to the simplicity of the eight dimensional manifold.
Our final goal is to reproduce the results found in this paper using more complicated CY fourfolds, in
which the D7 configurations include also intersecting branes. A first step would be to consider some
Voisin--Borcea manifold, modding out $K3\times K3$ by a freely acting involution. This breaks
the $SO(3)$ symmetry of $K3$ and gives a unique complex structure to the fourfold. Starting from such examples, we hope to further
develop our intuition for geometric moduli stabilisation in F-theory and eventually move forward to generic four-folds.

\vskip 2cm

\subsection*{Acknowledgements}

%\begin{center} \textbf{Acknowledgements} \end{center}

We are grateful to Hagen Triendl for discussions and to Rainer Ebert for comments on the manuscript.
This work was supported by SFB-Transregio 33  "The Dark Universe" by Deutsche
Forschungsgemeinschaft (DFG). CL acknowledges partial support from the
European Union 6th framework program MRTN-CT-2006-035863 "UniverseNet".

\clearpage
\appendix

\section{Lattice of Integral Cycles of \boldmath$K3$\label{sec:K3}}

The scalar product defined in~(\ref{K3modmetric}), or equivalently, the counting of oriented intersection
numbers of 2-cycles gives us a natural symmetric bilinear form on 
$H_2(K3,\mathbb{Z})$. It can be shown~\cite{a96} that with this scalar product, $H_2(K3,\mathbb{Z})$
is an even  self-dual lattice of signature $(3,19)$. By the classification of even self-dual
lattices we know that we may choose a basis for $H_2(K3,\mathbb{Z})$ such that the inner product
is characterised by the matrix
\begin{align}
\label{eq:UUUE8E8}
U \oplus U \oplus U \oplus \left(-E_8\right) \oplus \left(-E_8\right)\,,
\end{align}
where
\begin{align}
  U = \left(
    \begin{array}{cc}
      0 & 1 \\ 1 & 0 \\
    \end{array}
  \right) \ ,
\end{align}
and $E_8$ denotes the Cartan matrix of $E_8$.
% Choosing a point in the moduli space of $K3$ is
% now equivalent to choosing a time-like three-plane (i.e.~the metric is positive definite within the
% plane)  in $\mathbb{R}^{3,19}$  equipped with the inner product \eqref{eq:UUUE8E8}. This three-plane
% is spanned by the three vectors $\omega_i$ fulfilling the conditions~\eqref{eq:omegaconditions}. We
% identify $H_2(K3,\mathbb{R})$ and $H^2(K3,\mathbb{R})$ here and throughout this work.

Any vector in the lattice of integral cycles of an elliptically fibred $K3$ can now be written as
\begin{align} 
  D=p^{i}e^{i}+p_{j}e_{j}+q_{I}E_{I}\,,\label{H2}
\end{align}
where $i,j$ run from one to three and $I,J$ from $1$ to $16$. The $p_{i}$ as well as the $p^{i}$ are
all integers. The $E_{8}^{\oplus 2}$ lattice is spanned by $q_{I}$ fulfilling
$\sum_{I=1..8}q_{I}=2\mathbb{Z}$, $\sum_{I=9..16}q_{I}=2\mathbb{Z}$. In each of the two $E_{8}$
blocks, the coefficients furthermore have to be \emph{all} integer or \emph{all}
half-integer. The only nonvanishing inner products among the vectors in this expansion are 
\begin{align}
  E_{I}\cdot E_{J}&=-\delta_{IJ}\,,& e^{i}\cdot e_{j}&=\delta^{i}_{j} \,.
\end{align}

\vskip 1cm

\section{The Potential in Terms of \boldmath$W$ and $\check{W}$\label{VWWhat}}

\allowdisplaybreaks
For completeness, we also give the flux induced scalar potential in terms of two superpotentials. For a $CY_4$, it reads \cite{hl01}
\begin{align}
 V=\frac{e^K}{\Vv^3} \mathcal{G}^{\alpha\bar{\beta}} D_\alpha W D_{\bar{\beta}} \overline{W} +
 \frac{1}{\Vv^4}\left(\frac12 \check{\mathcal{G}}^{mn}\partial_m \check{W}\partial_n \check{W} -
   \check{W}^2\right) \,.
\end{align}
Here $K= -\ln \int_{CY_4} \Omega\wedge \overline{\Omega}$ and  $W$ and $\check{W}$ are given by
\begin{align}
 W&=\int_{CY_4} \Omega \wedge G_4\,,  & \check{W}&=\frac{1}{4} \int_{CY_4} J\wedge J \wedge G_4\,.
\end{align}
The complex structure moduli are labelled by $\alpha=1,\dotsc,h^{3,1}$, while $m=1,\dotsc,h^{1,1}$
counts the K\"ahler moduli.

For $K3\times K3$, we get a similar but not identical form. Note fist that the above
potential depends on $h^{1,1}+2h^{3,1}$ real moduli. This is the
dimension of the metric moduli space of a $CY_4$. But it is not the case for $K3\times \Kt$, whose
moduli space has dimension 
\begin{align}
 2 \times 58 = 2 \left( 3(h^{1,1}(K3)-1)+1\right)\,.
\end{align}
The moduli are the volume and the deformations of the $\omega_i$'s that are orthogonal to
all the $\omega_i$'s and whose number is then $h^2(K3)-3=h^{1,1}-1$. On the other hand ,
\begin{align}
  \begin{split}
    h^{1,1}\!\left(K3\times \Kt\right) +2h^{3,1}\!\left(K3\times \Kt\right)&\\ 
    &\mspace{-80mu}= 2 \left(\vphantom{\frac{a}{a}}h^{1,1}\!\left(K3\right) 
      +  2 h^{2,0}\!\left(K3\right)h^{1,1}\!\left(K3\right)\right) = 2\times  60\,. 
  \end{split}
\end{align}
This is again a reflection of the fact that for $K3$, only the three-plane itself is geometrically
meaningful: The two ``missing'' moduli correspond to the rotation of $j$ into real and imaginary
parts of $\omega$.

By an explicit computation one can get the new form of the potential:
\begin{align}\label{WWhpotential}
  \begin{split}
    V &= V_{G_{3,1}} + V_{G_{2,2}} \\
    &=\frac{e^{K}}{\Vv^3}  \mathcal{G}_{(0)}^{\alpha\bar{\beta}} D_\alpha W D_{\bar{\beta}} \overline{W}
      + \frac{1}{\Vv^4}\left(\frac12 \check{\mathcal{G}}^{mn}\partial_m \check{W}\partial_n \check{W}^2 -
        \check{W}^2\right)\,.
  \end{split}
\end{align}
The second term, $V_{G_{2,2}}$ is the same as for the $CY_4$ (note that
$m=1,...,h^{1,1}(K3)+h^{1,1}(\Kt)$). The only difference is in $V_{G_{3,1}}$: In the $CY_4$ case it
is given by the integral of  $G_{3,1}\wedge G_{1,3}$, where the subscript denotes the Hodge
decomposition. In that case it is also equal to the primitive part $G_{3,1}^{(0)}\wedge
G_{1,3}^{(0)}$, since  $G_{3,1}$ is automatically primitive. On $K3\times \Kt$, it is not primitive
and one must remove from $G_{3,1}$ the piece proportional to $J$. This is what the
metric $\mathcal{G}_{(0)}$ does. It is 
given by  
\begin{align}
\mathcal{G}_{(0)} = \left(\begin{array}{cc}
 -\frac{\int_{K3}\chi_\alpha\wedge \bar{\chi}_{\bar{\beta}}}{\int_{K3} \omega\wedge \bar{\omega}} & \\ &
	 -\frac{\int_{\Kt}\tilde{\chi}_\rho\wedge \bar{\tilde{\chi}}_{\bar{\sigma}}}{\int_{\Kt}
     \omegat\wedge \bar{\omegat}} 
\end{array}\right)\,,
\end{align}
where $\{\chi_\alpha\}$ is a basis for (1,1)-forms orthogonal to $\omega_3$.

The supersymmetry condition for the vacua can be written in terms of these two superpotentials. In
this case they assume the standard form (see for example \cite{gvw99,hl01,lmr05})
\begin{align}
 D_\alpha W &= 0\,, & W &= 0\,,& \partial_m \check{W} &= 0 \,.
\end{align}
The first two conditions say that the $G_4$ is a (2,2)-form, while the last one implies $G_4$ is
primitive.

\newpage
\section{Linear Algebra on Spaces with Indefinite Metric\label{app:linalg}}

Since some of the usual theorems about eigenvalues and eigenvectors of self-adjoint operators do not
carry over to the case of an indefinite scalar product, we collect some useful facts in this
appendix (see also \cite{indefMetric}). We consider a real vector space $\Vt$ equipped with a non-degenerate scalar product
$\left(\vt\cdot \wt\right)$ of signature $(n,m)$, where $n<m$ and $n$ refers to positive norm. In the
case we are interested in, $\Vt=H^2\!\left(\Kt\right)$ and the signature is $(3,19)$. Let $A$ be an
endomorphism of $\Vt$ which is selfadjoint with respect to this scalar product. We denote the set of
eigenvalues of $A$ by $\left\{\lambda_i\right\}$. Since the eigenvalues are the roots
of the real characteristic polynomial, they are either real or come in complex conjugate pairs. We
consider the complexification $\Vt_\mathbb{C}$ of $\Vt$, such that the scalar product involves complex
conjugation of the first entry.

In $\Vt_\mathbb{C}$, $A$ has $n+m$ eigenvalues. Note that a self-adjoint operator $A$ is not
necessarily diagonalisable in a space with indefinite metric.
However, this problem only occurs if there exists a zero-norm eigenvector
relative to a degenerate eigenvalue \cite{Pandit}. We will not consider this
non-generic case. Then $A$ is diagonalizable in $\Vt_\mathbb{C}$ with eigenvectors given by $\{e_i\}$.  From the selfadjointness, we have
\begin{align}\label{D1} 
  \left(\bar{\lambda}_i-\lambda_j\right)\left(e_i\cdot e_j\right)=0\,.
\end{align}
Since the metric is indefinite, $\left(e_i\cdot e_i\right)=0$
does not imply $e_i=0$, so that not all eigenvalues need to be real.

If there exist one non-real eigenvalue $\lambda$ with eigenvector $e$, then $\bar{\lambda}$ is also an eigenvalue.The corresponding eigenvector is $\bar{e}$. Equation \eqref{D1} tells us that $e$ and $\bar{e}$ are null. In the case we are considering, $\lambda$ is non-degenerate. Then, the non-degeneracy of the inner product implies $(\bar{e},e)\not =0$.
With these vector we can construct two real vectors
\begin{align}
  \vt^+&= e+ \bar{e}\,, & \vt^-&= -\I\left(e - \bar{e}\right)
\end{align}
that have opposite norm. Then, $\vt^\pm$ generate a subspace of the original real space $\Vt$, such that the scalar product on
this subspace is of signature $(1,1)$. One can define the orthogonal complement of this subspace in $\Vt$ and look for the next complex eigenvalue and the corresponding $2\times 2$ block. There can be at most $n$ of these $2\times 2$ blocks. Then there are at least $m-n$ real eigenvalues.

We conclude that the canonical form of a generic matrix $A$ selfadjoint with respect to a indefinite inner product with signature $(n,m)$ is block diagonal, with $n$ $2\times 2$ block relative to subspaces of signature $(1,1)$ and a positive definite $(m-n)$-diagonal block\footnote{
A matrix selfadjoint with respect to a definite metric is positive definite}. Vectors belonging to different blocks are orthogonal to each other.

Let us concentrate on a $2\times 2$ block. We choose a basis such that the metric has the matrix form
\begin{equation}
\Mt=\begin{pmatrix}
  0 &1\\  1& 0
\end{pmatrix}\:.
\end{equation}
The selfadjointness condition on $A$ is $A\,\Mt=\Mt\,A^T$, implying that
\begin{equation}
A=\begin{pmatrix}
  a &b\\  c& a
\end{pmatrix}\:.
\end{equation}
With a transformation that leaves $\Mt$ invariant, A can be brought to the canonical form\footnote{If $b,c$ are either both zero or both non-zero. Otherwise, the matrix is of the form we said before: It has a degenerate real eigenvalue relative to a zero norm eigenvector.}
\begin{equation}
A'=\begin{pmatrix}
  a &b\\  b& a
\end{pmatrix} \mbox{ or }
A'=\begin{pmatrix}
  a &-b\\  b& a
\end{pmatrix}\:.
\end{equation}
If we now change basis with the matrix $P=\frac{1}{\sqrt{2}}\begin{pmatrix} 1&1 \\ 1&-1  \end{pmatrix}$, then $M$ and $A$ go to:
\begin{equation}
\Mt=\begin{pmatrix}\label{Mt2x2}
  1 &0\\  0& -1
\end{pmatrix}\:, \qquad
A'=\begin{pmatrix}
   \lambda_1 &0\\  0& \lambda_2
\end{pmatrix} \mbox{ or }
A'=\begin{pmatrix}
  a &b\\  -b& a
\end{pmatrix}\:.
\end{equation}

\

Let us now specialise to the case of $A=G^a G$, i.e.~$V$ is another vector space, equipped with
a scalar product of the same signature, and $G$ is a map from $\Vt$ to $V$. $G^a$ denotes its adjoint
with respect to these scalar products, i.e.\ $\left(v,G \vt\right)=\left(G^a v,\vt\right)$ (where
$\vt\in \Vt$ and $v\in V$). Clearly, the composition $G^a G$ is a selfadjoint map from $\Vt$ to itself.

We want to determine the canonical form for $G$. It will be of the same structure of $A$, with $n$ $2\times 2$ blocks of signature $(1,1)$ and a diagonal part relative to a metric in the form $-{\mathbf 1}_{m-n}$. The diagonal part will be simply given by the square root of the diagonal block of $A$. 
Regarding the $2\times 2$ blocks, we find that both canonical forms can be written as $A'=g^a g$ with a ``square root'' matrix $g$.
%We take also for the corresponding vectors in $V$ a metric $M$ of the form \eqref{Mt2x2}. 
Since $A$ is of the form $G^aG$, the eigenvalues $\lambda_1,\lambda_2$ in \eqref{Mt2x2} must be either both positive or both negative. We consider these two cases separately. The canonical forms for $g$ are
\begin{align}
      g=\begin{pmatrix}
        \sqrt{\lambda_1} &0\\  0& \sqrt{\lambda_2}
      \end{pmatrix}\:,\qquad
      g=\begin{pmatrix}
        0& \sqrt{\left|\lambda_2\right|}\\ \sqrt{\left|\lambda_1\right|}& 0
      \end{pmatrix}\:,\quad
      g=\begin{pmatrix}
        \gamma & \delta\\-\delta & \gamma\\
      \end{pmatrix}\:,
\end{align}
where in the last matrix we have defined $\gamma$ and $\delta$ such that $\alpha=\gamma^2-\delta^2$ and $\beta=2\gamma\delta$.

Then, the matrix of $G$ can be brought with a change of basis into the form:
\begin{align}
  G_d&=
  \begin{pmatrix}
    g_1 &&&&&\\
    & \ddots&&&&\\
    & & g_n &&& \\
    &&& \sqrt{\lambda_1} &&\\
    &&&&\ddots  &\\
    &&&&&  \sqrt{\lambda_{n-m}}
  \end{pmatrix}\:.
\end{align}

If we call the matrix of the change of basis $\Pt$, then we can summarise our results as:
\begin{equation}
 \Pt^{-1} G^a G \Pt = G_d^aG_d \qquad,\qquad \Pt^T \Mt \Pt = \mathbb{M}
\end{equation}
where $\mathbb{M}$ is the diagonal matrix given by $n$ $2\times 2$ blocks $(+1,-1)$ and an $m-n$ block $(-1,...,-1)$.

\

We now show that there exists a change of basis in the space $V$ such that the matrix of $G$ can be brought to the form $G_d$, i.e. there exists a matrix $P$ such that 
\begin{equation}
P^{-1}G\Pt = G_d \:. 
\end{equation}
This matrix is given by $ P \equiv {G^a}^{-1} \Pt G_d^a$. Let us check that:
\begin{align}
 P^{-1} G \Pt = {G_d^a}^{-1}\Pt^{-1}G^aG\Pt = {G_d^a}^{-1}G_d^aG_d =G_d
\end{align}
Moreover, we obtain the relations:
\begin{equation}
 \Pt^{-1}G^a P = G_d^a \qquad\qquad P^{-1} GG^a P = G_dG_d^a \qquad\qquad P^T M P = \mathbb{M} \:.
\end{equation}

Only in the case of all eigenvalues being positive do we get a fully diagonal form for $G$, otherwise we
have non-diagonal $2\times2$ blocks.

Returning to the potential (and to the $K3$ case where $n=3$ and $m=19$), we see that if $G^aG$ is diagonalizable with non-negative eigenvalues, then $G$ and $G^a$ can be brought to the same diagonal form $G_d$ with respect to bases made up of three positive norm and nineteen negative norm vectors. This means that the minimum condition \eqref{eq:planetoplane} is satisfied. The converse is also true: If the condition \eqref{eq:planetoplane} is satisfied, then $G$ and $G^a$ can be brought to a diagonal form by changes of bases and so $G^aG$ becomes diagonal with non-negative entries.

\section{F-Theory Point in the K\"ahler Moduli Space\label{ftheorypoint}}

Let us fix two directions of the three-plane $\widetilde{\Sigma}$ to form the holomorphic two-form,
let us say $\omegat=\omegat_1+\I\omegat_2$, so $\jt=\sqrt{2\nu}\,\omegat_3$. We are left with 20
moduli: the 19 $\delta\omegat_2^m$ deformations of $\omegat_3$ and the volume $\nut$. 
These remaining 20 moduli can be parametrised with the 20 deformations of 
$\jt$ in $H^{1,1}(\widetilde{K3})$: 
\begin{align}\label{eq:jtagain}
  \jt = b \widetilde{B} + f \widetilde{F} + c^a \ut_a\,,  \qquad\qquad \mbox{with }
  \ut_a \mbox{ a basis }\bot \left<\widetilde{F},\widetilde{B},\omegat_1,\omegat_2\right> \,.
\end{align}
So we are essentially left with the K\"ahler moduli space.

The metric on this moduli space is ($i,j$ run over $\left\{b,f,c^a\right\}$)
\begin{align}\label{eq:modmetric}
 g_{ij} = -\partial_i\partial_j \log \left(\int \jt\wedge \jt\right) =-\partial_i\partial_j \log
 \left(2\,b(f-b)-c^a c^a\right)\,. %\qquad \mbox{where }\hat{t}^2=\sum_{a=1}^{18}
% (\hat{t}^a)^2 
\end{align}

We want to use this metric to compute the distance between one general point of the moduli space and
a point corresponding to the F-theory limit. As discussed in Section~\ref{sec:ftheory}, $b$ and $f$
give the volumes of fibre and base, and the F-theory limit involves $b\to 0$ while respecting the
bound~(\ref{eq:caconstraint}). We will consider a curve parameterised by $\epsilon$, 
\begin{align}\label{eq:modpath}
 b &= b_0\epsilon^2\,, &f &= \text{const.}\,, & c^a&=c^a_0 \epsilon\,,
\end{align}
where $c^a_0 c^a_0 = 2 \alpha b\left(f-b\right)$ and $\alpha\in\left[0,1\right)$ parameterises the
degree to which the bound is saturated. Note that the parameterisation~(\ref{eq:jtagain}) is simple,
but not exceedingly convenient. In particular, one might worry that the volume of $\Kt$ vanishes in
the limit of $\alpha\to 1$, even though base and fibre volume stay finite. However, before that
limit is reached, one can reparameterise the basis cycles such that the new $c^a$ are again
zero, while $f$ is now smaller than before. The limit $\alpha\to1$ is then the same as $\epsilon\to
0$.

The metric distance of the F-theory point from any other point ($\epsilon_0$) is given by $\int_{\epsilon_0}^0ds$, where
\begin{equation}
 ds = \sqrt{g_{ij} \dot{X}^i\dot{X}^j}\,d\epsilon \:.
\end{equation}
$X^i$ are $b,f,c^a$ and $\dot{X}^i$ are the derivatives of $X^i$ with respect to $\epsilon$.
% The metric~(\ref{eq:modmetric}) is not block-diagonal, so any explicit calculation becomes
% cumbersome. However, it is still straightforward to show that the F-theory point is infinitely far
% away: From the parameterisation~(\ref{eq:modpath}) we see that the volume is essentially
% proportional to $\epsilon^2$, 
% \begin{align}
%   \frac{1}{2} \jt\cdot\jt &= \epsilon^2 \left( b_0 \left(f-\epsilon^2 b_0\right)-\frac{1}{2} c^a_0
%     c^a_0\right)\,.
% \end{align}
By explicit calculation, one can show that all terms in the sum under the square root are of order $\epsilon^{-2}$ in the limit
$\epsilon\to 0$, times some finite coefficient. Hence, the metric distance from any finite point
$\epsilon_0$ to $\epsilon=0$ is 
\begin{align}
 \int^0_{\epsilon_0} \d s = \int^0_{\epsilon_0} \frac{\d\epsilon}{\epsilon} \cdot\left(\text{term
     finite for $\epsilon\to 0$}\right)\,,
\end{align}
i.e.~it diverges logarithmically.

\end{document}